\documentclass{emulateapj}

\newcommand{\Msol}{$M_{\odot}$}
\newcommand{\be}{\begin{equation}}
\newcommand{\ee}{\end{equation}}
\newcommand{\kms}{\mbox{km s$^{-1}$}}

\begin{document}


\title{The 10k zCOSMOS: morphological transformation of galaxies in the group environment since {\lowercase{$z$}} $\sim1$\footnotemark[1]}


\author{
K. Kova\v{c}\altaffilmark{2}, 
S. J.~Lilly\altaffilmark{2}, 
C. ~Knobel\altaffilmark{2}, 
M.~Bolzonella\altaffilmark{3}, 
A.~ Iovino\altaffilmark{4},
C. M.~Carollo\altaffilmark{2},
C.~ Scarlata\altaffilmark{5},
M.~ Sargent\altaffilmark{6},
O.~ Cucciati\altaffilmark{7},
G.~ Zamorani\altaffilmark{3},
L.~Pozzetti\altaffilmark{3},
L.~A.~M.~ Tasca\altaffilmark{7,8},
M.~ Scodeggio\altaffilmark{8},
P.~ Kampczyk\altaffilmark{2},
Y.~ Peng\altaffilmark{2},
P.~ Oesch\altaffilmark{2},
E.~ Zucca\altaffilmark{3},
A. ~Finoguenov\altaffilmark{9},
T.~Contini\altaffilmark{10},
J.-P.~ Kneib\altaffilmark{7},
O.~ Le F\`{e}vre\altaffilmark{7},
V.~Mainieri\altaffilmark{11},
A.~ Renzini\altaffilmark{12},
S.~Bardelli\altaffilmark{3},
A.~Bongiorno\altaffilmark{9},
K.~Caputi\altaffilmark{2},
G.~ Coppa\altaffilmark{3},
S.~ de la Torre\altaffilmark{7,4,8},
L.~ de Ravel\altaffilmark{7},
P.~ Franzetti\altaffilmark{8},
B.~ Garilli\altaffilmark{8},
F.~Lamareille\altaffilmark{10},
J.-F.~ Le Borgne\altaffilmark{10},
V.~ Le Brun\altaffilmark{7},
C.~Maier\altaffilmark{2},
M.~ Mignoli\altaffilmark{3},
R.~ Pello\altaffilmark{10},
E.~ Perez Montero\altaffilmark{10},
E.~ Ricciardelli\altaffilmark{13},
J.~D.~Silverman\altaffilmark{2},
M.~ Tanaka\altaffilmark{11},
L.~ Tresse\altaffilmark{7},
D.~ Vergani\altaffilmark{3},
U.~ Abbas\altaffilmark{7,14},
D.~ Bottini\altaffilmark{8},
A.~ Cappi\altaffilmark{3},
P.~ Cassata\altaffilmark{7,15},
A.~ Cimatti\altaffilmark{16},
M.~ Fumana\altaffilmark{8},
L.~ Guzzo\altaffilmark{4},
A.~M.~ Koekemoer\altaffilmark{17},
A.~ Leauthaud\altaffilmark{18},
D.~ Maccagni\altaffilmark{8},
C.~ Marinoni\altaffilmark{19},
H.~ J. McCracken\altaffilmark{20},
P.~ Memeo\altaffilmark{8},
B.~ Meneux\altaffilmark{9,21},
C.~ Porciani\altaffilmark{2,22},
R.~ Scaramella\altaffilmark{23},
N.~Z.~ Scoville\altaffilmark{24}
}

\footnotetext[1]{Based on observations
   obtained at the European Southern Observatory (ESO) Very Large
   Telescope (VLT), Paranal, Chile, as part of the Large Program
   175.A-0839 (the zCOSMOS Spectroscopic Redshift Survey)}

\altaffiltext{2}{Institute of Astronomy, ETH Z\"urich, CH-8093, Z\"urich, Switzerland; kovac@phys.ethz.ch}
\altaffiltext{3}{INAF Osservatorio Astronomico di Bologna, via Ranzani 1, I-40127, Bologna, Italy}
\altaffiltext{4}{INAF Osservatorio Astronomico di Brera, Milan, Italy}
\altaffiltext{5}{Spitzer Science Center, 314-6 Caltech, Pasadena, CA 91125, USA}\altaffiltext{6}{Max-Planck-Institut f\"{u}r Astronomie, K\"{o}ningstuhl 17, D-69117 Heidelberg, Germany}
\altaffiltext{7}{Laboratoire d'Astrophysique de Marseille, Marseille, France}
\altaffiltext{8}{INAF - IASF Milano, Milan, Italy}
\altaffiltext{9}{Max-Planck-Institut f\"ur extraterrestrische Physik,  
D-84571 Garching, Germany}
\altaffiltext{10}{ Laboratoire d'Astrophysique de Toulouse-Tarbes, Universite de Toulouse, CNRS, 14 avenue Edouard Belin, F-31400 Toulouse, France}
\altaffiltext{11}{European Southern Observatory, Karl-Schwarzschild- 
Strasse 2, Garching, D-85748, Germany}
\altaffiltext{12}{INAF - Osservatorio Astronomico di Padova, Padova, Italy}
\altaffiltext{13}{Dipartimento di Astronomia, Universita di Padova, Padova, Italy}
\altaffiltext{14}{INAF Osservatorio Astronomico di Torino, Strada Osservatorio 20, I-10025 Pino Torinese, Torino, Italy}
\altaffiltext{15}{Dept. of Astronomy, University of Massachusetts at Amherst}
\altaffiltext{16}{Dipartimento di Astronomia, Universit\'a di Bologna, via Ranzani 1, I-40127, Bologna, Italy}
\altaffiltext{17}{Space Telescope Science Institute, 3700 San Martin Drive, Baltimore, MD 21218}
\altaffiltext{18}{Physics Division, MS 50 R5004, Lawrence Berkeley National Laboratory, 1 Cyclotron Rd., Berkeley, CA 94720, USA}
\altaffiltext{19}{Centre de Physique Theorique, Marseille, Marseille, France}
\altaffiltext{20}{Institut d'Astrophysique de Paris, UMR 7095 CNRS, Universit\'e Pierre et Marie Curie, 98 bis Boulevard Arago, F-75014 Paris, France}
\altaffiltext{21}{Universit\"ats-Sternwarte, Scheinerstrasse 1, Munich D-81679, Germany}
\altaffiltext{22}{Argelander-Institut f\"{u}r Astronomie, Auf dem H\"{u}gel 71, D-53121 Bonn, Germany}
\altaffiltext{23}{INAF, Osservatorio di Roma, Monteporzio Catone  
(RM), Italy}
\altaffiltext{24}{California Institute of Technology, MS 105-24,
Pasadena, CA 91125, USA}


\begin{abstract}

We study  the evolution  of galaxies inside  and outside of  the group
environment since $z = 1$ using a large well defined set of groups and
galaxies from the zCOSMOS-bright  redshift survey in the COSMOS field.
The  fraction  of  galaxies  with  early-type  morphologies  increases
monotonically with  $M_B$ luminosity and stellar mass  and with cosmic
epoch. It is higher in  the groups than elsewhere, especially at later
epochs.  The  emerging environmental effect is superposed  on a strong
global   mass-driven   evolution,   and    at   $z   \sim   0.5$   and
$\log(M_*/M_{\odot}) \sim 10.2$, the  ``effect'' of group environment is
equivalent to (only)  about 0.2 dex in stellar mass or  2 Gyr in time.
The stellar mass function of galaxies in groups is enriched in massive
galaxies. We directly determine  the transformation rates from late to
early  morphologies,  and  for  transformations involving  colour  and
star-formation    indicators.     The    transformation   rates    are
systematically about twice as high in  the groups as outside, or up to
3-4  times higher  correcting for  infall  and the  appearance of  new
groups.  The rates reach values, for masses around the crossing mass $10^{10.5}$ \Msol, as high as 0.3 - 0.7 Gyr$^{-1}$ in the groups, implying transformation timescales of 1.4 - 3 Gyr, compared with less than 0.2 Gyr$^{-1}$, i.e. timescales $>$ 5 Gyr, outside of groups. 
All three transformation rates decrease at higher stellar masses, and 
must decrease also at the lower masses below $10^{10}$ \Msol\ which we cannot well probe. The rates involving colour and star-formation are consistently
higher  than those  for morphology,  by a  factor of  about  50\%. Our
conclusion is  that the transformations  which drive the  evolution of
the overall  galaxy population since $z  \sim 1$ must occur  at a rate
2-4  times higher  in groups  than outside  of them. 

\end{abstract}


\keywords{galaxies: clusters: general - galaxies: evolution - galaxies: high-redshift - galaxies: luminosity function, mass function - galaxies: structure}

\section{Introduction}

Galaxies exhibit a range of morphologies that reflect, at least in part, basic structural differences, such as the presence of
disks, spheroids and bars, as well as
differences in the visibility of the spiral arms due to for instance the rate of star-formation.
The commonly used morphological classification follows the Hubble sequence of elliptical, S0, spiral, barred spiral and irregular galaxies, with various subclasses \citep{Hubble.1926, Hubble.1936, Sandage.1961}. Galaxy morphology is undoubtedly a complex phenomenon reflecting different physical processes operating within and
around a given galaxy.  There are several underlying correlations involving morphology. At least up to $z \sim 1$, early type galaxies are observed to be redder, older, more luminous and more massive than spiral and irregular galaxies \citep[e.g.][]{Bell.etal.2004, Bundy.etal.2006, Pozzetti.etal.inprep, Bolzonella.etal.2009}. The fraction of rotationally supported early type galaxies is about $60\%$ and it does not change significantly between $z \sim 1$ and $z \sim 0$ \citep{vanderWel&vanderMarel.2008}. There are also systematic differences in the interstellar medium content. In the Local Universe, where HI systematic measurements are available, only a handful of elliptical galaxies have HI detected, against the plethora of spirals and irregulars, reflecting the much smaller $HI/M_*$ ratios in ellipticals compared with later type galaxies \citep[e.g.][]{Haynes&Giovanelli.1984, Noordermeer.etal.2005}.

It has been known for many years that the observed mix of different morphological classes at the present epoch also depends on the galaxian environment. Early type morphologies are preferentially found in dense, cluster-like regions. This idea can be traced back to Hubble, but the major impact on the astronomical  community started after Dressler's study of
55  nearby  galaxy  clusters \citep{Dressler.1980}.  The  existence of the morphology-density
relation is presumably linked to the physical processes which govern the
star formation (and morphological transformation) of galaxies and the connection of these 
to the environments of individual galaxies.

After  the  establishment    of    the   morphology-density    relation    in
clusters \citep[e.g.][]{Dressler.1980},  the   existence  of  the   relation  was
extended to the  group environments \citep[e.g.][]{Postman&Geller.1984} at low redshift. Later, the evolution with redshift was measured
\citep[e.g.][]{Dressler.etal.1997, Postman.etal.2005, Fasano.etal.2000, Desai.etal.2007}.  There is evidence that the evolution in
intermediate-density  environments has occurred  more recently than in
high-density environments \citep{Smith.etal.2005}. Also, it appears that the
morphological  mix  of  cluster  early-types  has  changed  with  time
\citep{Fasano.etal.2000, Postman.etal.2005}. Ultimately, understanding the origin of the morphology-density relation can help to 
disentangle the relative roles of nature (conditions at epoch of galaxy formation) and  nurture (environment dependent processes) in determining the 
properties of galaxies.

It is known that galactic mass also plays a major role in determining the properties of
galaxies, since both colour, specific star-formation rates and morphologies are
strongly correlated with the stellar mass of galaxies \citep[e.g.][]{Kauffmann.etal.2003}.
The recent evidence that the stellar mass function of galaxies itself depends on
the environment \citep{Baldry.etal.2006, Bolzonella.etal.2009} complicates the
separation of the relative roles of mass (which could be regarded as
``nature'') and of environment (``nurture''). This linkage also complicates the
interpretation of environmental effects in any sample of galaxies
that contains a significant range of masses, since different environments
will contain, within such a sample, a different mix of masses. This in turn
can easily produce the appearance of a ``spurious'' environmental dependence through the dependence of galaxian properties on mass 
(see \citealt{Tasca.etal.2009} and \citealt{Cucciati.etal.submitteda} for 
a discussion).

Several physical processes have been suggested which could enhance the transformation of late to early type galaxies (and star-forming to quiescent) in dense regions (see \citealt{Boselli&Gavazzi.2006} for a recent review of these processes). Mergers and interactions with other galaxies or with the cluster potential \citep[e.g.][]{Toomre&Toomre.1972, Barnes.1988} are purely gravitational processes. Galaxy harrasment combines the cumulative effect of the interaction of galaxy with close neighbours with the interaction of the total cluster potential \citep{Moore.etal.1998}. Ram pressure stripping may be responsible for the removal of the cold gas when a galaxy infalls with high velocity into a dense medium \citep{Gunn&Gott.1972, Quilis.etal.2000}. While ram-pressure stripping is expected to be most effective in the dense intracluster medium, \citet{Rasmussen.etal.2006} have shown that gas stripping by the intragroup medium may also be rapid enough to transform late types to S0s over a few Gyrs. Strangulation (or starvation) is a process that may remove a halo of hot diffuse gas after a galaxy becomes a satellite in a larger dark matter halo \citep{Larson.etal.1980, Balogh&Morris.2000}. \citet{Bamford.etal.2009} summarised these processes in to two broad categories: (a) those processes which will quench star formation and act on morphology passively, through the fading of spiral arms and reduction of the prominence of the disc after star formation has stopped (e.g. strangulation), and (b) those processes which will directly affect the stellar kinematics and galactic structure and also lead to a cessation of star formation (e.g. mergers). While both categories can produce S0s, only the second one can produce true elliptical galaxies.

One way to assess the relevance of a type-transformation process is to constrain its timescale. However, there are number of timescales related to a given process, and the interpretation of an observed transformation rate is not necessarily unique.  We here differentiate between the following timescales. The ``physical timescale $t_p$'' is a measure for a duration of the transformation of an individual galaxy. For example, the $t_p$ of the morphological transformation would be the average time which elapses since the beginning of the morphological transformation, when a galaxy would still be classified as late, till the point in its evolution when it will be classified as early. Sometimes it is more relevant to express the transformation in a probabilistic sense, as in radioactive decay. We can then quantify how long it will take (on average) for a galaxy of a given class before it gets transformed to a galaxy of some other class. This is a ``statistical timescale'' $t_s$, which is just the inverse of the average transformation rate. Finally, there is an observed ``rate timescale'' $t_{\eta}$ (obtained as the inverse of the observed rate), based on the real observations, whose meaning depends on its relation to the other timescales. If $t_p \ll t_{\eta}$, then $t_{\eta}$ will be of order $t_s$. However, if $t_p$ is of order $t_{\eta}$ then $t_{\eta}$ will be more a measure of $t_p$.  The interpretation of the observed $t_{\eta}$ therefore depends on the physical timescale $t_p$.

In this paper, the main aim is to study the changes in the morphological mix of
galaxies in group environments that have occurred in the time interval between $z \sim 1$ down to $z \sim 0.1$. Our primary goal is to understand the role of the group environment in the transformation of galaxies over this redshift range. We carry out the analysis by measuring the redshift evolution of the fraction of galaxies with early type morphologies in the total population of galaxies of a given luminosity or stellar mass in three different environmental samples: group, field and isolated galaxies. This leads us to construct the morphological transformation rate, normalised to the number of galaxies in the ``pre-transformation'' state (i.e. late morphological type) in different environments and quantify the timescales of the morphological transformation in different environments.  We then compare these rates with the equivalent ones defined by overall colour and by star-formation rate indicators, that reflect the quenching of star-formation activity. Determination of the transformation rates, i.e. of the flux of galaxies crossing through a chosen dividing line, should be less sensitive to the exact values chosen to divide galaxies into different classes than for example the quantity which reflects directly the partitioning into these classes of galaxies (such as crossing or transition mass defined as the mass at which two classes are equally partitioned).

The analysis is based on the COSMOS field \citep{Scoville.etal.2007a}, relying heavily on the zCOSMOS data \citep{Lilly.etal.2007, Lilly.etal.2009} necessary for the precise, small-scale definition of environment on the scale of groups. This paper is closely related to the number of other studies based on the data in the COSMOS field that have been aimed to address the complex interplay between galaxy properties and environment up to $z \sim 1 - 1.5$. A complementary analysis exploring galaxy colours in the group catalogue of \citet{Knobel.etal.2009} is presented in \citet{Iovino.etal.submitted}. \citet{Scoville.etal.2007b, Capak.etal.2007a, Guzzo.etal.2007, Cassata.etal.2007} and \citet{Ideue.etal.2009} study the galaxy property-environment relation using the datasets based on the photometric redshifts. \citet{Tasca.etal.2009, Cucciati.etal.submitteda} and \citet{Bolzonella.etal.2009} extend this study by using high precision spectroscopic redshifts and the continuous density field of \citet{Kovac.etal.2009}. Moreover, the environmental dependence of specific galaxy populations has been addressed in separate studies, e.g. infrared-luminous galaxies in \citet{Caputi.etal.2009}, AGNs in \citet{Silverman.etal.2009} and post-starburst galaxies in \citet{Vergani.etal.submitted}. The relation between the galaxy distribution and that of the underlying matter density field is explored in \citet{Kovac.etal.submitted}.  

Throughout the paper we use a concordance cosmology with $\Omega_m=0.25$, $\Omega_{\Lambda}=0.75$ and $H_0=70$ \kms Mpc$^{-1}$. Magnitudes are given in the AB system.

\section{The zCOSMOS survey}

\subsection{Details of the survey}

COSMOS \citet{Scoville.etal.2007a} is a multi-wavelength survey of a 2 square degree equatorial
field, observed so far by  the major earth and space based facilities,
e.g. Hubble,  Spitzer,   Galex,  Chandra,   Subaru.   The  zCOSMOS 
\citep{Lilly.etal.2007} is
a spectroscopic redshift survey in this field carried out with the VIMOS  spectrograph 
on the ESO UT3 8-m VLT.  The
zCOSMOS-bright part of the survey is flux limited at $I_{AB} < 22.5$, and will ultimately
obtain over 20,000 spectra. At this flux limit, the redshifts of the majority
of the observed galaxies fall in  the range $0<z<1.4$. A higher
redshift  range, $1.4<z<3$ is targeted in  the zCOSMOS-deep programme.
Here, we use the  first 10,000 spectra from the zCOSMOS-bright, which we
refer to as the ``10k sample'' \citep{Lilly.etal.2009}.  The high-quality  photometry in the COSMOS field also 
allows us  to derive accurate photometric
redshifts,  with an uncertainty of $0.023(1+z)$ or better. These are
used to understand the spectroscopic completeness of the survey \citep[see][]{Lilly.etal.2009}.

The rest-frame absolute magnitudes of all $I_{AB}<22.5$ targets are measured using the spectral energy distribution (SED) fitting employing the ZEBRA code \citep{Feldman.etal.2006}. The magnitudes are obtained as the best fit normalised to each galaxy photometry \citep{Capak.etal.2007b} at the best available redshift, spectroscopic or photometric (\citealt{Oesch.etal.inprep}, see also \citealt{Zucca.etal.submitted}). Stellar masses $M_*$ are obtained from  fitting stellar  population
synthesis models to the SED of the observed magnitudes as described in \citet{Bolzonella.etal.2009} and \citet{Pozzetti.etal.inprep}, using the best available redshift. Stellar masses which we use are obtained with the
\citet{Bruzual&Charlot.2003} libraries and the Chabrier  initial mass
function \citep{Chabrier.2003}, the \citet{Calzetti.etal.2000} extinction law with $0 < A_V < 3$ and  solar metallicities. The assumed star  formation history (SFH) is  decreasing  exponentially  with a time  scale  $\tau$, taking values from the interval $0.1<\tau<30$ Gyr. The stellar  mass is  obtained by
integrating the  SFH over  the galaxy age, from which  the mass  of gas  processed by
stars   and  returned   to  the   interstellar  medium   during  their
evolution  (``return fraction'') has been subtracted. See \citet{Bolzonella.etal.2009} for more details.

\subsection{The zCOSMOS 10k group catalogue}

The
details of our group-finding algorithm and the group catalogue
is presented in  \citet{Knobel.etal.2009}. Here, we outline only
the most important points regarding the group catalogue. 

\citet{Knobel.etal.2009} combine two standard group searching algorithms: Friends-Of-Friends (FoF) and the 
Voronoi-Delaunay method (VDM) to define groups in the 10k zCOSMOS. The group finding parameters are optimised based on the tests on the COSMOS mock catalogues \citep{Kitzbichler&White.2007} designed to match the geometrical and sampling properties of the 10k zCOSMOS catalogue. \citet{Knobel.etal.2009} also developed a new ``multi-pass'' strategy that optimises group finding over a range of richnesses by first optimising for the largest groups and then subsequently moving down to groups with a smaller number of observed members.

\begin{figure*}
\centering
\includegraphics[width=0.45\linewidth]{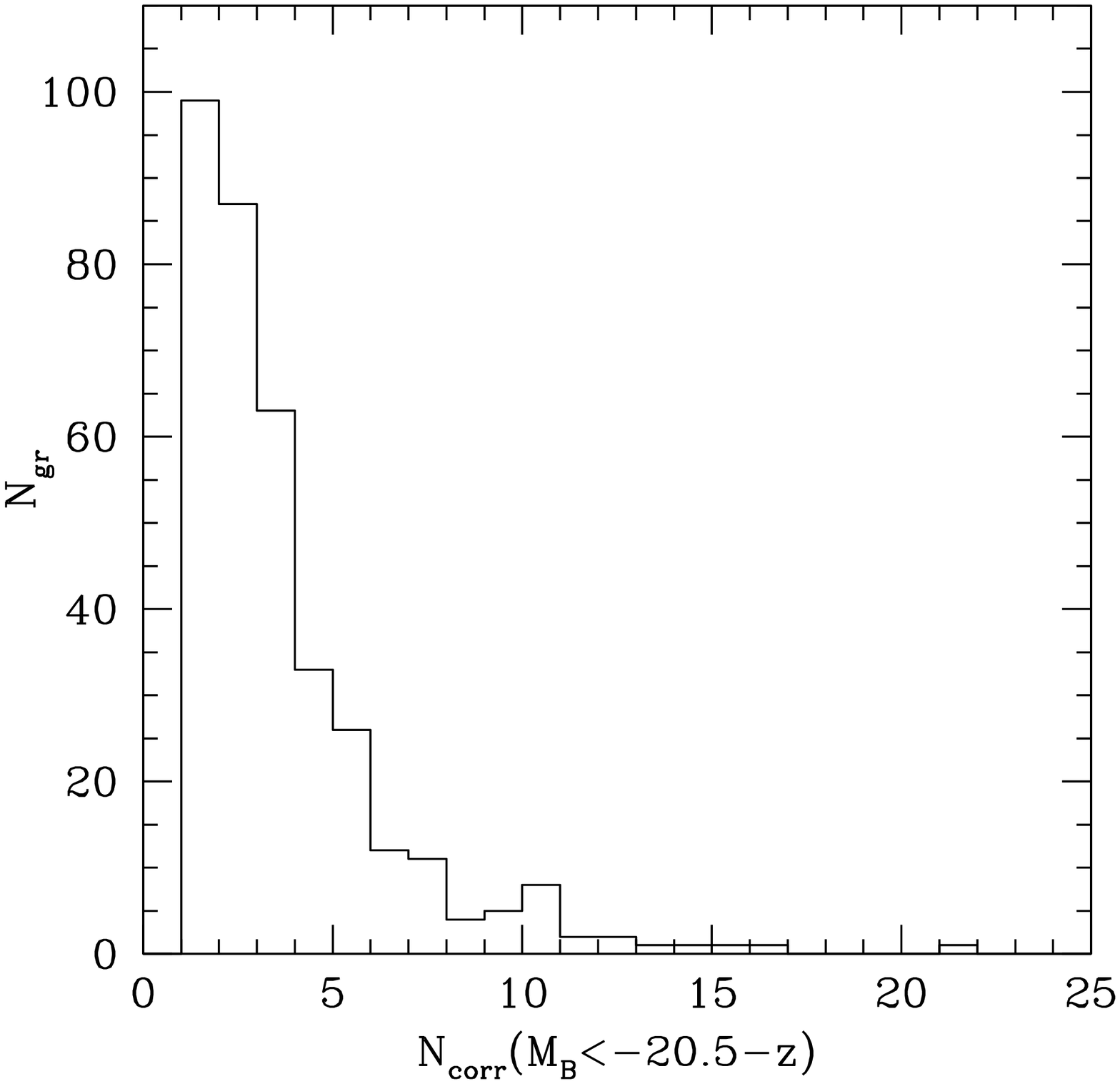}
\includegraphics[width=0.45\linewidth]{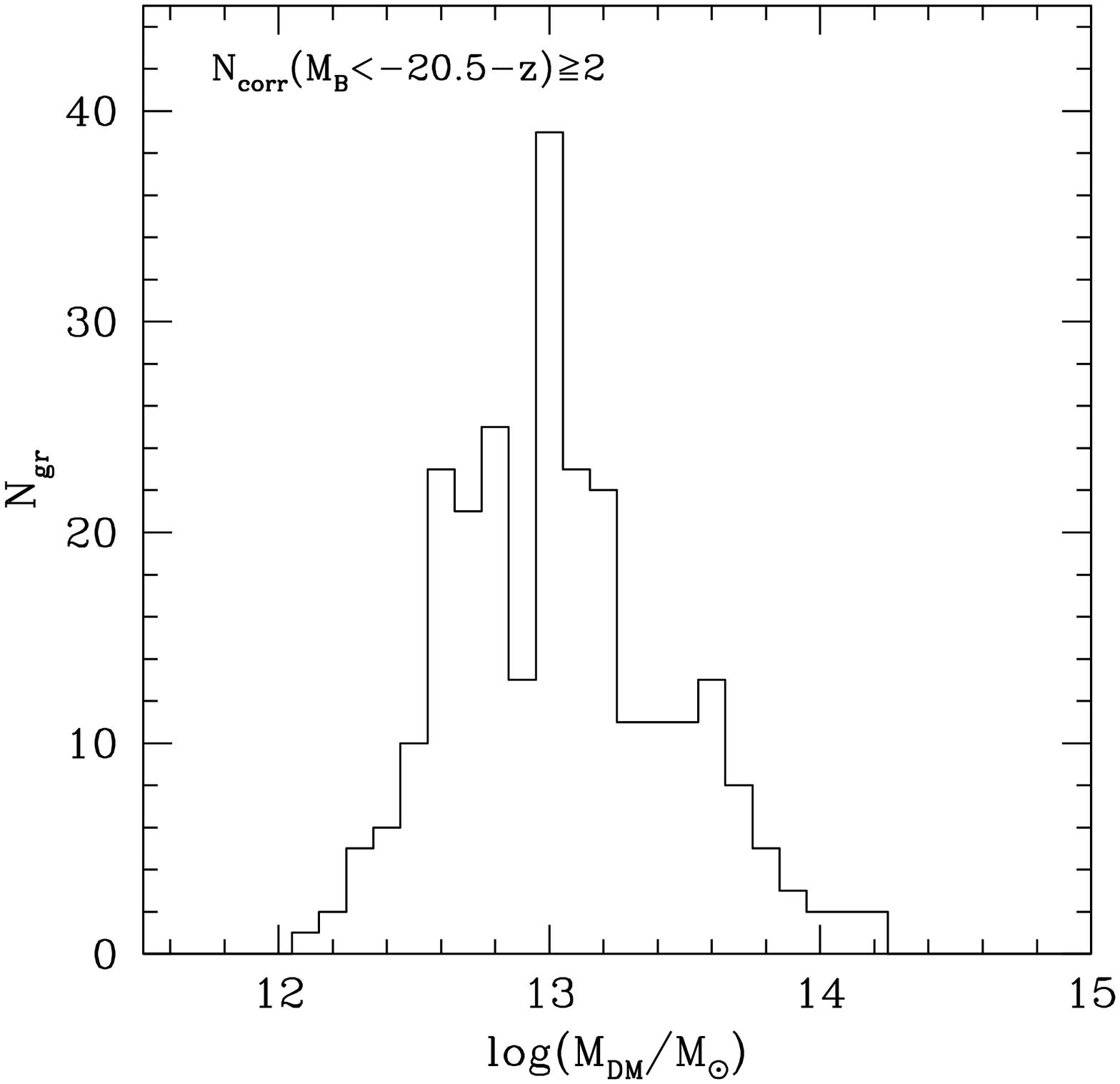}
\caption{Properties of the zCOSMOS groups ($0.1<z<1$) used for the current analysis. Left: Distribution of the ``corrected richness'', i.e. the number of members corrected for sampling incompleteness, above $M_B<-20.5-z$. Right: Distribution of the estimated halo masses for the zCOSMOS groups shown on the left. The halo masses are estimated by comparison with the COSMOS mock catalogues of \citet{Kitzbichler&White.2007}. The majority of the groups have less than 10 corrected $M_B<-20.5-z$ members and masses in the range $12<\log(M_{DM}/M_{\odot})<14$, emphasizing that the sample involves cosmic structures well below the masses of rich clusters of galaxies.}
\label{fig_groupmass}
\end{figure*}

The group catalogue which we use in this work is the so-called ``one-way-matched''  catalogue, which is the intersection of the  independently created FoF and VDM catalogues. In total, there are 800 groups with at least two members in the current ``10k-sample''. Over 100 of these groups have at least 5 spectroscopically observed members. 

Extensive comparisons with the mock catalogues \citep{Kitzbichler&White.2007} has established that the zCOSMOS 10k group catalogue has a relatively high purity and completeness - 82$\%$ and 81$\%$ respectively for groups with $N \ge 5$, and only marginally lower for the poorer groups. It should however be appreciated that these numbers are based on comparisons with the ideal group catalogues that could have been constructed (with two or more spectroscopic members per group) from a sparsely sampled
``10k-like'' sample, rather than the actual set of all groups in the COSMOS volume. 
As a result, some of the galaxies that are not currently placed in a group are nevertheless likely to be 
in a group which will be identified in the future with the higher sampling of the full survey. The current ``non-group'' sample will therefore contain some ``group'' galaxies, and this contamination will act to reduce observed differences between group and non-group populations.  Partly for this reason, as discussed below, we will usually use the entire galaxy population that we have observed spectroscopically, selected without regard to environment, as our control sample. This also reflects
the traditional definition of a ``field sample''.

Due  to the non-uniform sampling  scheme  of the  current 10k zCOSMOS  survey
\citep{Lilly.etal.2007, Lilly.etal.2009}, the fraction of group members observed (as well as the chance of recognising the poorer groups) will be a function of
positions on the sky.  We therefore correct the observed
number of members for each group by the overall sampling fraction of the survey at that location, taking into
account both the spatially variable spectroscopic sampling rate and the 
redshift-dependent success rate in yielding a high confidence redshift (using the corrections derived in \citealt{Knobel.etal.2009}). We will refer to this as the corrected richness, which is the estimated true number of members
above the flux-limit of the survey.   For any given physical group, this corrected richness will vary
with redshift due to the changing luminosity limit that is associated with the observed $I$-band 
limit.   We can deal with this by considering groups with a given corrected richness above some fixed luminosity-limit, allowing this limit to evolve with redshift to account, as best we can, 
for the luminosity evolution of individual galaxies.  We refer to such a set of groups as a ``volume-limited'' sample in that it should contain all groups of a given type, but still this sample could be potentially affected by the so called ``progenitor bias'' since infall of new members will produce new groups in this volume-limited sample at later epochs. Moreover, given that the original group catalogue is generated from a flux limited sample, there will be a slight bias with redshift because poor groups are more difficult to detect at high $z$ than at low $z$. However, the distribution of corrected richness with redshift is rather flat, reassuring us that this effect is minor.

The distribution of corrected richnesses using the luminosity limit $M_B<-20.5-z$ in $0.1<z<1$ is shown in the left panel in Figure~\ref{fig_groupmass}. The majority of the 10k zCOSMOS groups have less than 10 members (corrected for sampling) above $M_B<-20.5-z$.  

\citet{Knobel.etal.2009} also give estimates for the observed velocity dispersion for groups with $N \ge 5$. \citet{Knobel.etal.2009} also give an estimate of the virial mass of the dark matter (DM) halo for all of the groups by assigning a halo mass to each group using the empirical relation between the halo mass and the corrected observed richness, calibrated at each redshift on the mock catalogues \citep{Kitzbichler&White.2007}. The distribution of these estimated halo masses for the zCOSMOS groups with corrected richness of at least two members with $M_B<-20.5-z$ detected in $0.1<z<1$ is shown in the right panel in Figure~\ref{fig_groupmass}. The groups which we are using for the analysis typically have halo masses in the range $12<\log(M_{DM}/M_{\odot})<14$. This emphasises that the zCOSMOS volume to $z \sim 1$ does not contain rich clusters of galaxies.

\subsection{Isolated galaxies}
\label{sec_isolated}

In parallel, we measure the morphological mix of galaxies in a sample of isolated galaxies, to further highlight the possible environmental segregation of galaxy morphological types. We use for this purpose the sample of isolated galaxies defined by \citet{Iovino.etal.submitted} as follows.  First, the Voronoi volume is estimated for each galaxy in the zCOSMOS flux limited $I_{AB}<22.5$ sample in the area defined by $149.57<RA<150.41$ deg and $1.76<DEC<2.68$ deg in $0.1<z<1$. These Voronoi volumes are normalised by the median Voronoi volume in $\Delta z = 0.2$ centred at each galaxy redshift, to correct for the changing mean-intergalaxy separation with redshift. Galaxies with the highest quartile of the normalised Voronoi volume are defined as the isolated galaxies. Moreover, galaxies with too large Voronoi volumes with respect to the normalised mean Voronoi volume ($\sim10\%$ of the preliminary sample of isolated galaxies), most probably affected by the survey borders, and galaxies detected to be in groups ($\sim14\%$ of the preliminary sample of isolated galaxies) are removed from the sample.  Based on the $I_{AB}<22.5$ mock catalogues, about 90$\%$ of the galaxies defined as isolated using the described procedure should be the single occupants of their haloes.

\subsection{Field galaxies}

Following original practice, we refer in this paper to a ``field'' sample as being all galaxies selected without regard of environment.  In the comparisons of the various relations, the field sample will always be chosen to match exactly the selection of the group and isolated samples of galaxies as regards all non-environmental parameters, e.g. luminosity, mass, type, redshift etc.

\subsection{Morphologies}
\label{sec_morphdef}

The COSMOS  field is  the largest contiguous  field covered by HST
ACS images. The HST observations \citep{Koekemoer.etal.2007} were taken in the  F814W filter.
The ACS imaging allows the possibility to derive robust structural
parameters  and to confidently morphologically  classify all  galaxies into 'Hubble types'  down to about 24 mag (and possibly deeper), well below the flux limit of zCOSMOS-bright.

Here we use  the structural parameters and morphological classification based on the {\it Zurich Estimator of Structural Types (ZEST)} and presented in \citet{Scarlata.etal.2007}.
The ZEST classification scheme is  based on a
principal component analysis of five non-parametric diagnostics of
galaxy structure:  asymmetry A,  concentration C, Gini  coefficient G,
the second-order moment of the  brightest 20$\%$ of galaxy pixels M$_{20}$
and  the ellipticity $\epsilon$.  The morphological  classification is
performed in the space of  the first three eigenvectors, which contain
most    of    the   variance    of    the   original    non-parametric
quantities.   At each point in the three-dimensional eigenvector space,
a morphological class is assigned based on eye-ball inspection and the median
Sersic index of all the galaxies in that cell.

Galaxies  are classified by ZEST  into ellipticals, disk and irregular galaxies.
The  disk  galaxies are  further  split in  four
bins,  namely bulgeless, small-bulge, intermediate-bulge and  bulge-dominated
disk galaxies.

In the following we  consider as  ``early-types'' all galaxies that are classified 
by ZEST as either 'elliptical' or 'bulge-dominated' systems, i.e. specifically ZEST Classes 1 and 2.0. All other ZEST types are together considered to be  ``late-type'' galaxies.   In terms of the conventional Hubble classification scheme, our division
would probably be between Sa and Sb.

%%%%%%%%%%%%%%%%%%%%%%%%%%%%%%%%%%%%%%%%%%%%%%%%%%%%%%%

\section{Sample definition}
\label{sec_sampledef}

While there are about 40,000 galaxies in the zCOSMOS field which satisfy
the  survey  selection criteria  $I_{AB} <  22.5$  mag (the ``40k sample''
catalogue), at  the current stage  of the project only about 10,000 spectra
(10k  sample  catalogue)  have  been  obtained.  In  this sample  every
redshift has been  assigned a quality flag based  on the confidence of
the  measured  spectroscopic  redshift. This quality flag is calibrated using 
both the repeatability of
the spectroscopic measurement and the consistency of the photometric redshift estimates (see \citealt{Lilly.etal.2007} and \citealt{Lilly.etal.2009}
for  more details). 
In this paper we use all galaxies with spectroscopic
confidence class 3 and 4 and all galaxies with a lower spectroscopic
confidence class but whose spectroscopic consistent with their rather accurate photometric redshift. 
The sample consists of 8478 galaxies, representing overall
88$\%$  of  the  10k  zCOSMOS-bright  sample at $z<1.2$, increasing to 95$\%$ for $0.5 < z < 0.8$, 
with an overall redshift reliability of 98.6$\%$. We refer to this sample as the high
confidence class 10k sample (HCC 10k).

Due to the inhomogeneous sampling of galaxies with spectroscopic redshifts, we limit our analysis to the 
zCOSMOS region defined by $149.55 \le RA \le 150.42$ deg and $1.75 \le DEC \le 2.70$ deg, in which the sampling is both higher and more homogeneous. The area is slightly smaller for the isolated galaxies, see Section~\ref{sec_isolated}). This is the same region as used by Iovino et al. (submitted) for a parallel analysis which investigate the colour segregation of galaxies in groups (e.g. see Figure 1 in Iovino et al. submitted). 

\subsection{Correction for the spectroscopic incompleteness}

In zCOSMOS, the targets to be observed spectroscopically are selected independently of the
galaxy properties, except for the requirement of $15.0 < I_{AB} < 22.5$. 
The failure to assign a redshift to an observed
spectra may have several causes, some of which may depend directly or indirectly on 
the properties of the target galaxies.  We  therefore adopt a 
weighting scheme  to correct for the objects observed spectroscopically which
have insufficiently good redshifts to be included in  the HCC 10k sample (including
complete  failures) taking  into  account only  galaxy properties.  We
define  a  multidimensional  space  by the selection  magnitude  $I_{AB}$, the
rest-frame  $(U-V)_0$ colour, the morphological type  and the redshift, relying on the values calculated using  the high quality photometric redshifts.  For each galaxy,
the multi-parameter space is sampled for galaxies of the same morphological class within an interval, centred on the galaxy in question,
that is 0.25 mag in $I_{AB}$  magnitude, 0.25 mag
in the rest-frame $(U-V)_0$ colour, and 0.1 in redshift.
We then define the inverse weight $w_i$ of each galaxy $i$ in the HCC 10k sample as the fraction:

\be  w_i = \frac{N_{HCC10k}^i}{N_{10k}^i},
\label{eq_complweights} \ee

\noindent 
where $N_{HCC10k}^i$ is the number of galaxies in the HCC
10k sample and $N_{10k}^i$ is the number of galaxies 
in the total  10k sample which are neighbours of  the $i$-th galaxy in
the     above    defined     manner     in    the     multidimensional
$I_{AB}-(U-V)_0-z_{phot}$-morphology space. In those cases where a galaxy
in  the  10k  sample could not be assigned any  of  the  properties  considered
($I_{AB}, (U-V)_0,  z_{phot}$ and morphology) the  galaxy is excluded
from the  analysis. We  can neglect these galaxies  given that their exclusion is based on
the absence of data and not directly to the properties of galaxies. Moreover, 
they constitute a negligible fraction ($< 4\%$) of the total sample. 
The final number of the group, field and isolated galaxies is 1859, 
5671 and 1147 in $0.1<z<1$ and in the selected area (see above), respectively.

\subsection{Computing the fraction of galaxies and errors}

Having defined the inverse weights of the galaxies in the HCC 10k sample 
using equation~\ref{eq_complweights},  the  fraction of
galaxies of  a given  morphological type within any given galaxy sample is
straightforwardly computed from the HCC sample in terms of the sum of 
these individual weights. Specifically, the total number of galaxies of a 
given morphological type $T$ (using the ZEST morphological classification) in a given
sample of $k$ galaxies is calculated using

\be  N_{T}^K = \Sigma_k \frac{1}{w_k(T)} \ee

\noindent
where the summation goes over all galaxies of the $T$ type in the K-subsample of the HCC 10k sample. 
Similarly, the total (weighted) number of galaxies of any type can be calculated using 

\be N_{ALL}^K = \Sigma_k \frac{1}{w_k} \ee
\noindent
summing over all galaxies in the K-subsample of the HCC 10k sample. 
The fraction of galaxies of a given morphological type $T$ is then straightforwardly given by 

\be  f_{T}^K = \frac{N_{T}^K}{N_{ALL}^K}.  \ee

\noindent
The index $K$ refers to the sample in which the morphological
fraction of galaxies is calculated and is defined by environment, as well luminosity or stellar mass.  
In considering sets of galaxies defined by luminosity or mass, it was decided to only consider intervals
in either quantity in which we are essentially complete over a given redshift range, i.e. not to attempt a $V/V_{max}$ 
type correction for incompleteness.

Uncertainties in the observed fractions are estimated using a bootstrap resampling of the given subsample of galaxies, 
setting the error bars in the $f_T^K$ as the standard deviation in the distribution of fractions resulting from 1000 Monte Carlo bootstrap realisations.

%%%%%%%%%%%%%%%%%%%%%%%%%%%%%%%%%%%%%%%%%%%%%%%%%%%%%%%%%%%%%%%%

\section{Results}

In this section, we measure and compare the morphological mix of galaxies 
residing in different environments (groups, isolated and field) lying within the zCOSMOS volume in
the  redshift interval  $0.1 <  z <  1$.  As noted above, we refer to the morphological 
types 1 and 2.0 together as the ``early type'' galaxies and the remainder as ``late type''.

Given that the zCOSMOS groups contain typically only a few members, we cannot measure a meaningful morphological fraction of galaxies in individual groups, as is common when studying similar relations in rich clusters of galaxies, but rather combine all galaxies residing in groups with similar properties to build a statistically meaningful sample of galaxies.  It will become clear that, even with this large set of high redshift groups, the selection of sub-samples of groups and galaxies for study is primarily guided by the need to have statistically usable numbers of galaxies in each of the bins.

Furthermore, the current sparse sampling of zCOSMOS means that we cannot meaningfully distinguish between central and 
satellite galaxies within the groups.

Much recent work aimed to address the redshift evolution of a morphological mix of galaxies targeted rich clusters individually which could be easily identified at high $z$ (see recent work by Desai et al. 2007 and references therein). For the precise reconstruction of the groups one needs high resolution spectroscopic redshifts in a relatively large continuous volume. Needless to say, the additional requirement is high resolution imaging over the full survey area in order to reliably measure galaxian morphologies. The zCOSMOS groups are identified in a larger volume than in the two previous high redshift ($z \sim 1$) spectroscopic surveys, VVDS \citep{LeFevre.etal.2005} and DEEP2 \citep{Davis.etal.2003}. Moreover, only colour properties of galaxies in group environments have been studied in $0.25<z<1.2$ \citep{Cucciati.etal.submittedb} and in $0.75<z<1.3$ \citep{Gerke.etal.2007} in the VVDS and DEEP2, respectively.

\subsection{Morphological fraction of galaxies in the  luminosity complete samples}

\begin{figure}[t]
\includegraphics[width=0.95\linewidth]{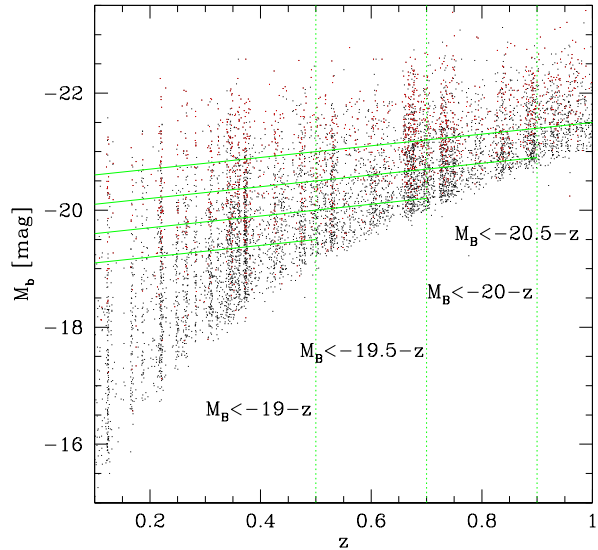}
\caption{The rest frame $B$-band magnitude as a function of redshift for the 10k zCOSMOS sample. Galaxies of early morphological type (ZEST types 1 and 2.0) are coloured red. The green lines define the four luminosity-complete samples used in the analysis.}
\label{fig_mBz}
\end{figure}

\begin{table*}
\centering
\begin{tabular}{c|cccc}
\hline
$M_{B,compl} \ limit$ & $M_B<-19-z$  & $M_B<-19.5-z$ & $M_B<-20-z$& $M_B<-20.5-z$ \\
$z \ range$ & $0.1<z<0.5$ & $0.1<z<0.7$ & $0.1<z<0.9$ & $0.1<z<1$ \\
\hline
$Group (M_B<M_{B,compl})$ & 564 &   665 &  605 &  329 \\
$Group (M_B<20.5-z)$ & 275 & 321 & 386 & 329 \\
$Field$ & 1391 & 1924 & 2098 & 1319 \\
$Isolated$ & 240 & 334 & 340 & 206 \\
\hline
\end{tabular}
%\end{center}
\caption{\label{tab_lum}Number of (non-weighted) galaxies in the luminosity complete samples.}
\end{table*}

To study the dependence of morphological segregation of galaxies in groups on the luminosity of galaxies, it is useful to first define "luminosity-complete", i.e. ``volume-limited'', samples of galaxies.  We do this primarily to compare with the results in literature, before urning to mass-selected samples below. 

We define four galaxy samples which satisfy the following  criteria: $M_B < -19-z$ at $0.1 <
z < 0.5$, $M_B < -19.5-z$ at $0.1  < z < 0.7$, $M_B < -20-z$ at $0.1 <
z < 0.9$ and $M_B < -20.5-z$  at $0.1 < z < 1$. We chose to use the (rest-frame) $M_B$ magnitude because it is well matched to the observed $I$-band selection of zCOSMOS-bright at high redshifts, with an exact match at $z \sim 0.8$, thereby assuring that the completeness of the zCOSMOS galaxies is not colour dependent \citep[see][]{Iovino.etal.submitted}. The ``-z'' term accounts, at best approximately, for
the luminosity evolution of individual galaxies $\Delta M_B = -1 \times \Delta z$ (see \citealt{Kovac.etal.2009} for discussion). The number of galaxies in the luminosity complete samples is given in Table~\ref{tab_lum}. As shown in 
Figure~\ref{fig_mBz}, these galaxy samples should be statistically complete.

Correspondingly, for each of these luminosity-complete galaxy samples, we can also define a set of ``luminosity-complete'' groups that satisfy the group criteria, i.e. that their corrected richness is at least two above these same galaxy 
luminosity limits.  These groups should also be ``volume-limited'' in a sense, although it is clear 
that the accretion of new members may lead to a change in the group population through the relevant redshift 
range.

\begin{figure*}[t]
\centering
\includegraphics[width=0.9\linewidth]{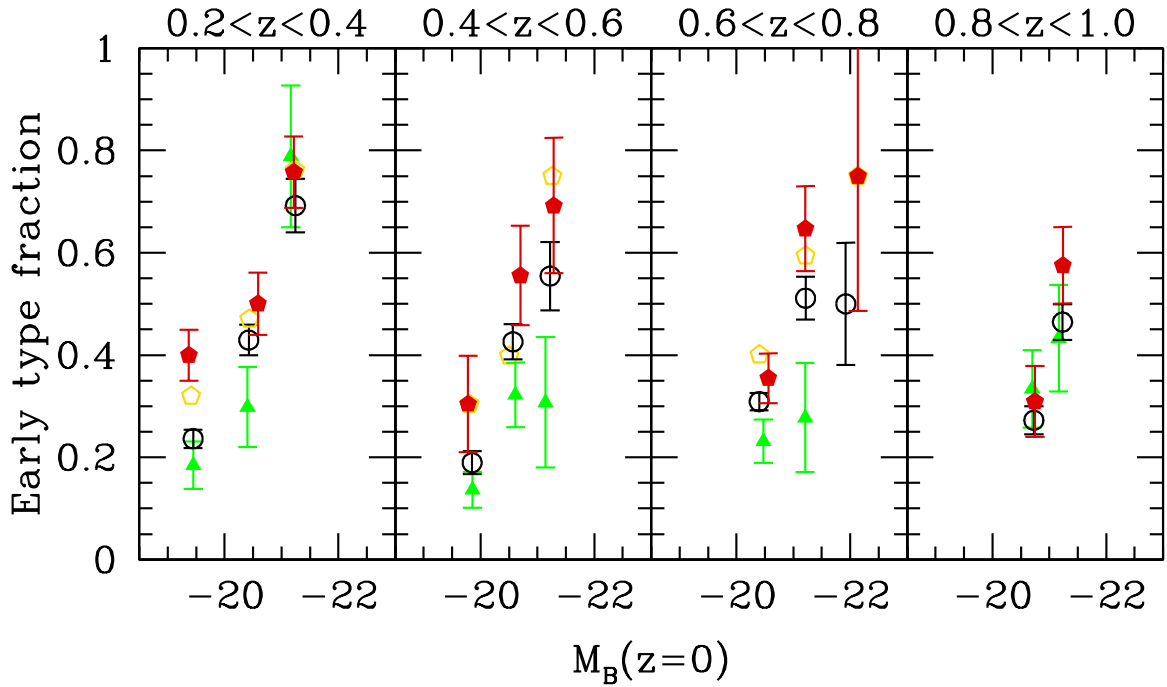}
\caption{The fraction of early type galaxies at a given $M_B$ magnitude, evolving the luminosity to $z=0$. The red and yellow pentagons represent the group galaxies, the black circles represent the field galaxies and the green triangles represent the isolated galaxies. The group galaxies are included in the analysis only if the group has a corrected richness of at least two in the given luminosity complete sample (empty yellow pentagons) or if the group has the effective richness of at least two in $M_B<-20.5-z$ sample (solid red pentagons, which are therefore complete across all panels). The fraction of galaxies in the different panels are calculated using only those galaxies from luminosity complete samples in the narrow redshift slices (indicated at the top of each panel). The fraction of early type galaxies is higher for brighter galaxies in all redshift bins. The fractions of early type galaxies at fixed $M_B$ are also higher for galaxies residing in groups compared with the overall population of field galaxies. The fractions of isolated galaxies with early type morphologies tend to be the lowest.}
\label{fig_morphfraclum}
\end{figure*}

With these considerations and caveats in mind, Figure~\ref{fig_morphfraclum} shows the fraction of early type galaxies 
in different environments as a function of (evolving) absolute $M_B$ magnitude. 
The black points show the field sample (all galaxies) while the green points show the set of isolated galaxies. The yellow points in each panel show the subset of galaxies that appear in the corresponding complete
set of groups at that redshift, i.e. with more than two corrected members above a luminosity threshold for which we are
complete, i.e.  $M_B < -19-z$ for the $0.2 < z < 0.4$ panel, $M_B < -19.5-z$ for the $0.4 < z < 0.6$ panel,
$M_B < -20-z$ for the $0.6 < z < 0.8$ panel. The red points represent galaxies in groups with two or more
members above $M_B < -20.5-z$, for which we are complete across the whole redshift range.  Apart from the caveats above about infall and progenitor bias, the
group galaxies that are represented by the red points should therefore be directly comparable across the whole redshift range.

It is clear from Figure~\ref{fig_morphfraclum} that the fraction of early type galaxies is higher for brighter galaxies in all redshift bins, covering the range from $z=0.1$ to $z=1$.  This reflects the strong and well-known dependence of morphology on mass or luminosity.  In addition, however, the fraction of early type galaxies at fixed $M_B$ is higher for galaxies residing in groups compared with the overall population of field galaxies. This holds for every luminosity sample and at every redshift bin considered, even though the difference is sometimes within the errors.  Similarly, the fractions of isolated galaxies with early type morphologies tend to be lower than in the sample as a whole.  In the highest redshift bin ($0.8<z<1$), the differentiation of the early type fraction of galaxies in the different environments becomes negligible. Here, the morphological mix of early type galaxies is within the errors the same in all environments.

\begin{figure*}
\centering
\includegraphics[width=0.9\linewidth]{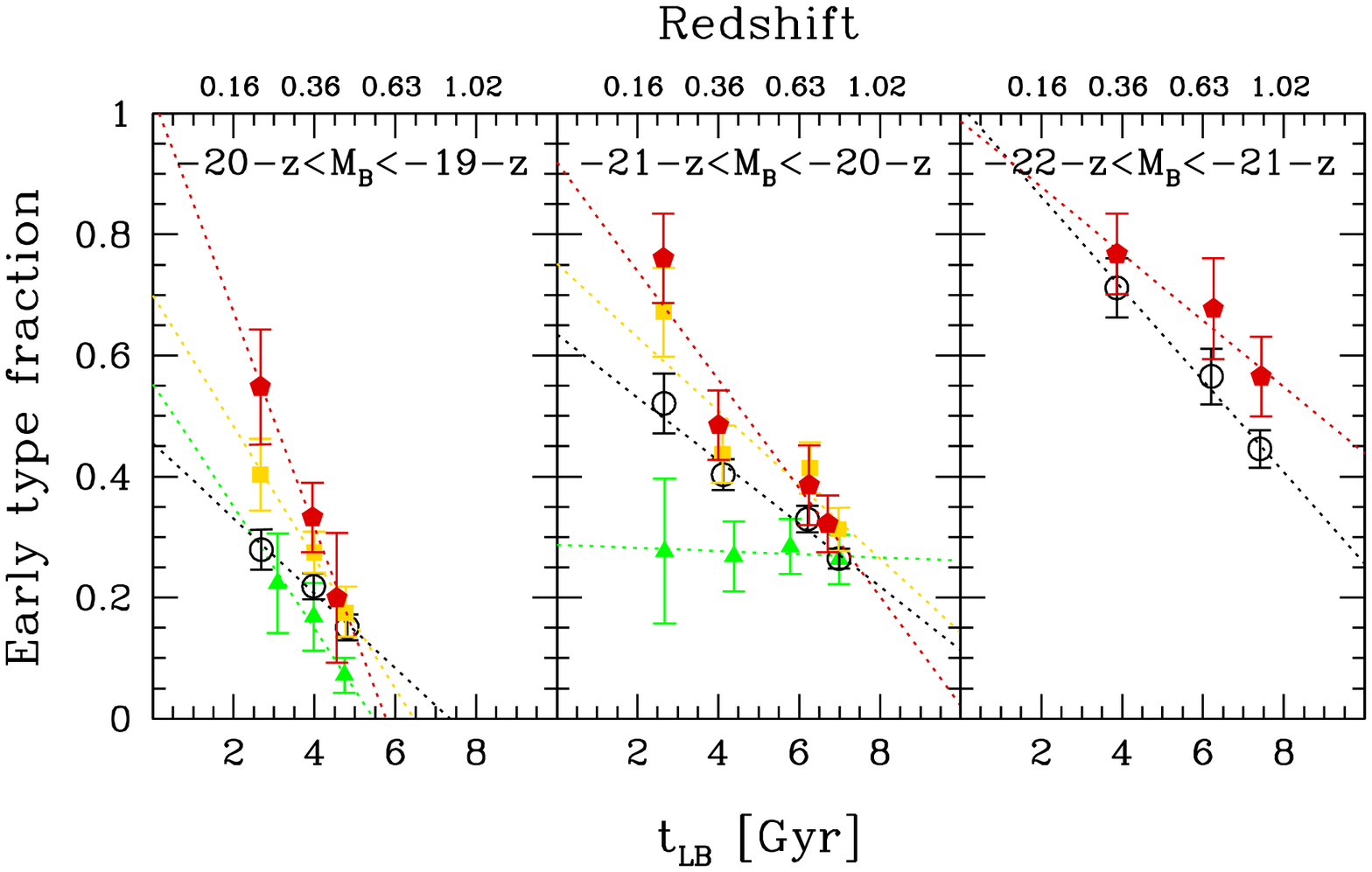}
\caption{The redshift/look-back time evolution of the fraction of early type galaxies in different environments. As in Figure 3, the red and yellow symbols represent the group galaxies, black circles represent the field galaxies and the green triangles represent the isolated galaxies in the indicated $M_B$-bins of galaxies. We consider two following types of groups for this analysis: with the corrected richness of at least two in $M_B<-19-z$ (yellow symbols left panel), in $M_B<-20-z$ (yellow symbols middle panel) or in $M_B<-20.5-z$ sample of galaxies (red symbols, therefore complete across all panels). For the analysis, we consider only redshift intervals in which galaxies of a given luminosity are drawn from the luminosity complete samples: $0.1<z<0.5$, $0.1<z<0.9$ and $0.2<z<1$ going from the left to the right, respectively. The symbols are plotted along the horizontal axis at the median look-back time/redshift of galaxies in the considered bin. The results indicate a build up over time of the population of early type galaxies in both the group environment and the field environment (all galaxies) and, especially at fainter $M_B$, also in the isolated sample.}
\label{figure_morphfractime_lum}
\end{figure*}

Comparing the panels across the diagram, it is clear that the fraction of early type galaxies at a given luminosity increases with time.  This is more clearly seen in Figure~\ref{figure_morphfractime_lum}, which shows the redshift/look-back time evolution of the fraction of early type galaxies in three narrow ranges of $M_B$ which are explored over three redshift intervals: $0.1<z<0.5$, $0.1<z<0.9$ and $0.2<z<1$. As in Figure~\ref{fig_morphfraclum}, the red points 
represent the members of a set of groups that is, in principle, volume-limited up to $z=1$, whereas the yellow points represent the members of 
a set of groups that is only volume limited within each individual panel. The redshift evolution of the early type fraction seen in Figure~\ref{figure_morphfractime_lum} indicates a clear build up over time of the population of luminous early type galaxies in the group environment. For the isolated galaxies the build-up of early type galaxies is detected only for the faintest galaxies. For the isolated galaxies with $-21-z<M_B<-20-z$ the fraction of early types stays more or less constant since $z=0.9$, but the error on the measured fraction at the lowest redshift is rather large and we do not have a sufficient number of isolated galaxies with $-22-z<M_B<-21-z$ to reliably carry out this analysis. 

As in Figure~\ref{fig_morphfraclum}, it can be seen that the fraction  of  early type  galaxies in the 
group environment is higher than the fraction
of early type galaxies in the field (all galaxies), which is again higher than the early-type fraction 
in the isolated galaxies. By fitting a simple linear relation between the fraction of early type galaxies and the look-back time, the data suggest that the morphological transformation of galaxies from late to early type has generally happened earlier for galaxies residing in groups than for galaxies of the same luminosity in the field, coupled with an overall trend for the transformation to occur earlier for more luminous galaxies.  We return to quantify this further below.

The evolution of the early type fraction over cosmic time and its dependence on luminosity qualitatively mirrors (i.e. it is inverse to) the evolution and its environmental and luminosity dependencies of the fraction of blue galaxies in the zCOSMOS group, field and isolated samples, investigated by \citet{Iovino.etal.submitted}. 

\subsubsection{Comparison with literature}

An exact comparison of our results with the previously published studies of the morphological mix of galaxies in $0.1<z<1$ is hindered by the different morphological classifications, and further 
complicated by the different selection of galaxies and the generally much denser environments. However, we can qualitatively place our results within the previous results obtained with the luminosity complete samples.

Galaxies which reside in the zCOSMOS groups reside also in the regions of the highest density in the continuously defined zCOSMOS density field \citep{Kovac.etal.2009}. The higher fractions of the early type galaxies in the groups relative to the field and isolated (when measured) samples are consistent with the morphology-density relations of  \citep{Tasca.etal.2009, Capak.etal.2007a, Guzzo.etal.2007, Cassata.etal.2007} at similar luminosity limits and similar investigated environments.

The ZENS project \citep{Carollo.etal.inprep} is a $B$ and $I$ wide-field imaging survey of galaxies in a statistically and luminosity-complete sample of 185 groups at $z \sim 0.05$ extracted from the 2PIGG catalogue \citep{Eke.etal.2004} with more than five catalogued galaxy members, in order to measure morphologies and quantify substructure (e.g. bars, bulges and disks), and determine the presence and properties of faint structures. The observed increase of the fraction of early type galaxies with luminosity in the group environment is clearly detected in the $z \sim 0.05$ ZENS groups down to $M_B=-17$ \citep{Cibinel.etal.inprep}. Going to higher redshift, \citet{Wilman.etal.2009} measured the morphological mix of $M_r<-19$ galaxies in the group and non-group environments using a sample of optically selected groups in $0.3 \le z \le 0.55$ from the Canadian Network for Observational Cosmology redshift survey \citep{Carlberg.etal.2001, Wilman.etal.2005}. They conclude that the fraction of S0 galaxies in groups at $z \sim 0.4$ is as at least as high as in clusters and X-ray selected groups, and it is higher than the fraction of S0s of the sample of galaxies not classified to belong to a group. The fraction of elliptical galaxies at fixed luminosity does not appear to be significantly different in the group and non-group environment. Taking the fractions of S0s and ellipticals together, the results of \citet{Wilman.etal.2009} and our results at $z \sim 0.4$ are in qualitative agreement, in the sense that at this redshift we measure a higher fraction of early type galaxies in the groups than in the field or isolated (i.e. non-group) sample of galaxies at all luminosities.

\citet{Desai.etal.2007} combined the  data from the various studies of the morphological content in clusters in order to construct the redshift evolution of the morphological mix of galaxies. Using the fractions of early type galaxies (ellipticals and S0s) measured in 10 ESO Distant Cluster Survey \citep[EDisCS; ][]{White.etal.2005} clusters in $0.5<z<0.8$ \citep{Desai.etal.2007} and the literature data compiled by \citet{Fasano.etal.2000}, both down to $M_V=-20$ (using cosmology with $\Omega_m=1$ and $H_0=50$ \kms Mpc$^{-1}$), \citet{Desai.etal.2007} conclude that the fraction of S0 galaxies in clusters increases about 2 times from $z=0.4$ to $z=0$, while the fraction of ellipticals stays constant over the same redshift interval. The EdisCS results ($M_V<-20$) combined with the equivalent studies of seven clusters from \citet{Postman.etal.2005} at $0.8<z<1.27$ ($M_V<-19.27-0.8z$) are consistent with no evolution in the morphological mix in clusters over the entire $0.4<z<1.25$ interval (using $\Omega_m=0.3$, $\Omega_{\Lambda}=0.7$ and $H_0=70$ \kms  Mpc$^{-1}$, \citealt{Desai.etal.2007}), suggesting that $z \sim 0.4$ is an epoch at which the S0 fraction in cluster cores begins to grow. However, our results are consistent with the constant build-up of early types since $z=1$ or $z=0.9$ in at least group and field environments for $-22-z<M_B<-21-z$ or $-21-z<M_B<-20-z$ galaxies, respectively. Ignoring the differences in the luminosity selection (the median $B-V$ colour of the field $M_B<-20.5-z$ or $M_V<-20$ zCOSMOS galaxies in $0.1<z<1$ is 0.53 or 0.46, respectively), the different redshift evolution of early type galaxies in the clusters and groups indicate that in $0.4<z<1$, groups are preferred  environment for the  morphological transformation of galaxies over clusters.

\subsection{Evolution of the morphological fraction as a function of the properties of the group}

In the previous Section we have seen the changes in the morphological mix of galaxies in groups over cosmic time.  Comparison of the yellow and red points
in the panels on the left of Figure~\ref{fig_morphfraclum} and Figure~\ref{figure_morphfractime_lum} suggests a dependence also on the group properties, since the yellow points represent groups that extend to poorer levels than the red points, which represent galaxies in (rich) groups that have at least two members above 
$M_B<-20.5-z$. \citet{Desai.etal.2007} have shown that morphological evolution of galaxies at $z<1$ is stronger in lower mass clusters than in the most massive ones and there have been number of claims that the strongest evolution between $z \sim 1$ and $z \sim 0$ in the galaxy properties has been in the intermediate mass \citep[e.g.][]{Poggianti.etal.2006} or density \citep[e.g.][]{Smith.etal.2005} environments. 

We can use the  group corrected richness and measured velocity dispersion to study the morphological segregation as a function of a group property in our own sample. While the best way to isolate the effect of a group property would be to use galaxies in a narrow range of physical properties selected in a narrow bin of group properties, that type of analysis is not possible with the current 10k zCOSMOS sample due to the small number statistics. Therefore, we use instead the luminosity complete samples of galaxies in groups of similar properties.

\begin{figure}
\centering
\includegraphics[width=0.98\linewidth]{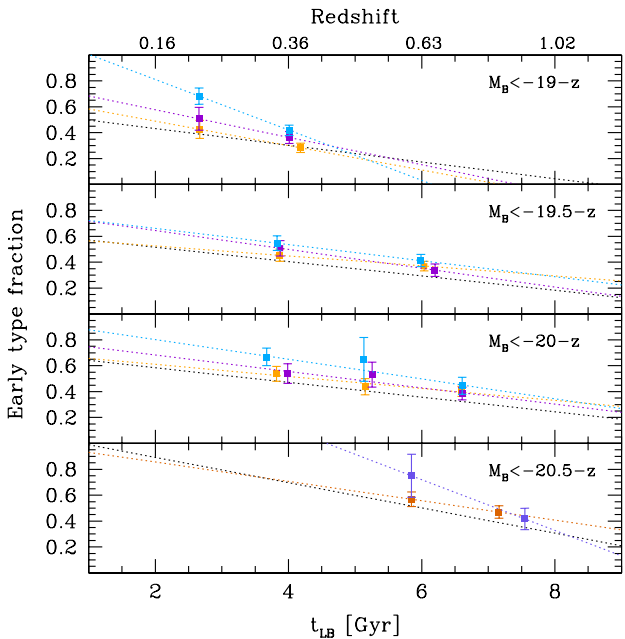}
\caption{The time-evolution of the fraction of early type galaxies in the groups of various richnesses. The solid squares represent the group population in luminosity complete samples as indicated in the individual panels, which are also used to define the corrected richness. The orange, violet and blue squares are for the galaxies residing in the groups with effective richness between 2 and 5, between 5 and 8 and larger than 8, respectively, for the top three panels. In the bottom panel, the light red and violet-blue squares are for the groups with the effective richness between 2 and 7, and larger than 7, respectively. The points are plotted at the median look-back time of galaxies in that bin. The fits are carried out at the median look-back time in the given redshift bin. For a comparison, the continuous line represents the linear fit to the $f_{early}-t_{LB}$ relation of the field galaxies in the corresponding luminosity complete sample. At lower luminosities and/or lower redshifts there is a clear trend of  higher early-type fraction in the richer groups.}  
\label{fig_morphfracneff}
\end{figure}

\subsubsection{Group richness}

First, we look at the dependence of the morphological fraction of galaxies as a
function of the corrected richness of the host group. We carry out this study for the four luminosity complete samples defined above, using the group corrected richness defined in each of the luminosity complete samples, i.e. the corrected number of members above the given galaxian luminosity limit. We split the group galaxies in bins using the effective richness of a group (above the quoted luminosity limit) between 2 and 5, between 5 and 8 and larger than 8 for the groups in the luminosity complete samples up to the luminosity of $M_B<-20-z$. For the brightest sample $M_B<-20.5-z$ we define bins of group effective richness between 2 and 7 and larger than 7, considering only galaxies and groups in $0.4<z<1$ in order to keep the statistics meaningful. The results are shown in  Figure~\ref{fig_morphfracneff}.    There is a clear trend, especially for lower luminosity galaxies, for the richest groups to have a higher early-type fraction, especially as the redshift decreases and for lower luminosity galaxies.

\subsubsection{Velocity dispersion}

\begin{figure}
\centering
\includegraphics[width=0.98\linewidth]{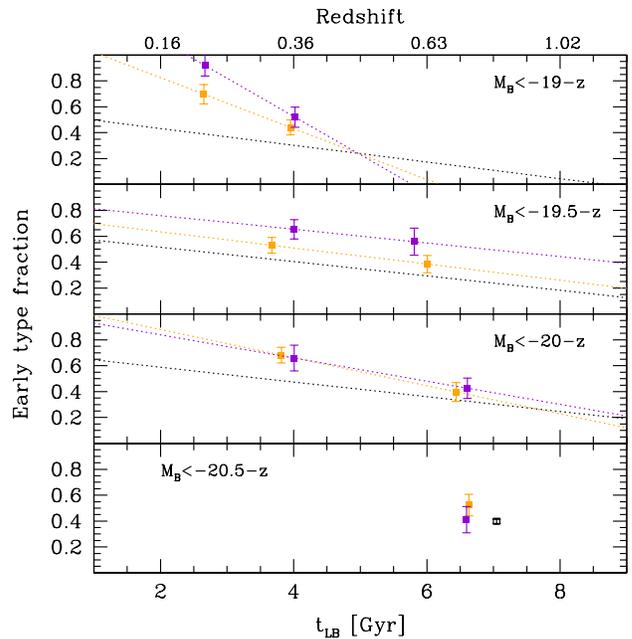}
\caption{The fraction of early type galaxies in different environments defined by the velocity dispersion of a group, as a function of redshift/time. The yellow and violet solid squares are for the galaxies residing in the groups with the velocity dispersion between 250 and 500 \kms, and larger than 500 \kms, respectively. The additional requirement is that a group has at least 5 observed ($I_{AB} < 22.5$) members. In the given panels, all galaxies used for the analysis satisfy the indicated magnitude limit, and the redshift ranges in which these samples are complete.  For a comparison, the continuous line represents the linear fit to the $f_{early}-t_{LB}$ relation of the field galaxies in the corresponding luminosity complete sample. The bottom panel is an exception where we consider a sample of $M_B<-20.5-z$ galaxies only in $0.6<z<1$ in order to keep the number of galaxies statistically meaningful. For a comparison, we plot the fraction of early type field galaxies measured in the same redshift range, marked with the black open symbol. The fraction of early type galaxies is larger in the groups with larger velocity dispersion for galaxies fainter than $M_B=-19.50-z$.}
\label{fig_morphfracvsigma}
\end{figure}

This analysis can be repeated using the observed velocity dispersion for the groups. Due to the large uncertainties related to velocity dispersion 
measurements \citep[see][]{Knobel.etal.2009}, only  groups  with at
least 5 observed members (above the $I_{AB} < 22.5$ cut) are used for this analysis. Using the four previously defined luminosity complete samples of galaxies, we divide the groups into two sets according to their velocity dispersion: $250<\sigma<500$ \kms and $\sigma>500$ \kms. The fraction of early type galaxies in these two group bins are shown in Figure~\ref{fig_morphfracvsigma}. 

Even though of a lower significance, the same picture emerges.  The fraction of early type galaxies is larger in the groups with larger velocity dispersion for galaxies fainter than $M_B=-19.5-z$. For brighter galaxies we do not detect any significant difference between fractions of early type galaxies in groups of different velocity dispersion, possibly due to the limited number of galaxies.

\subsection{Importance of the stellar mass distributions (stellar mass function in the groups and for the isolated galaxies)}
\label{sec_gsmf}

\begin{figure*}
\centering
\includegraphics[width=0.485\linewidth]{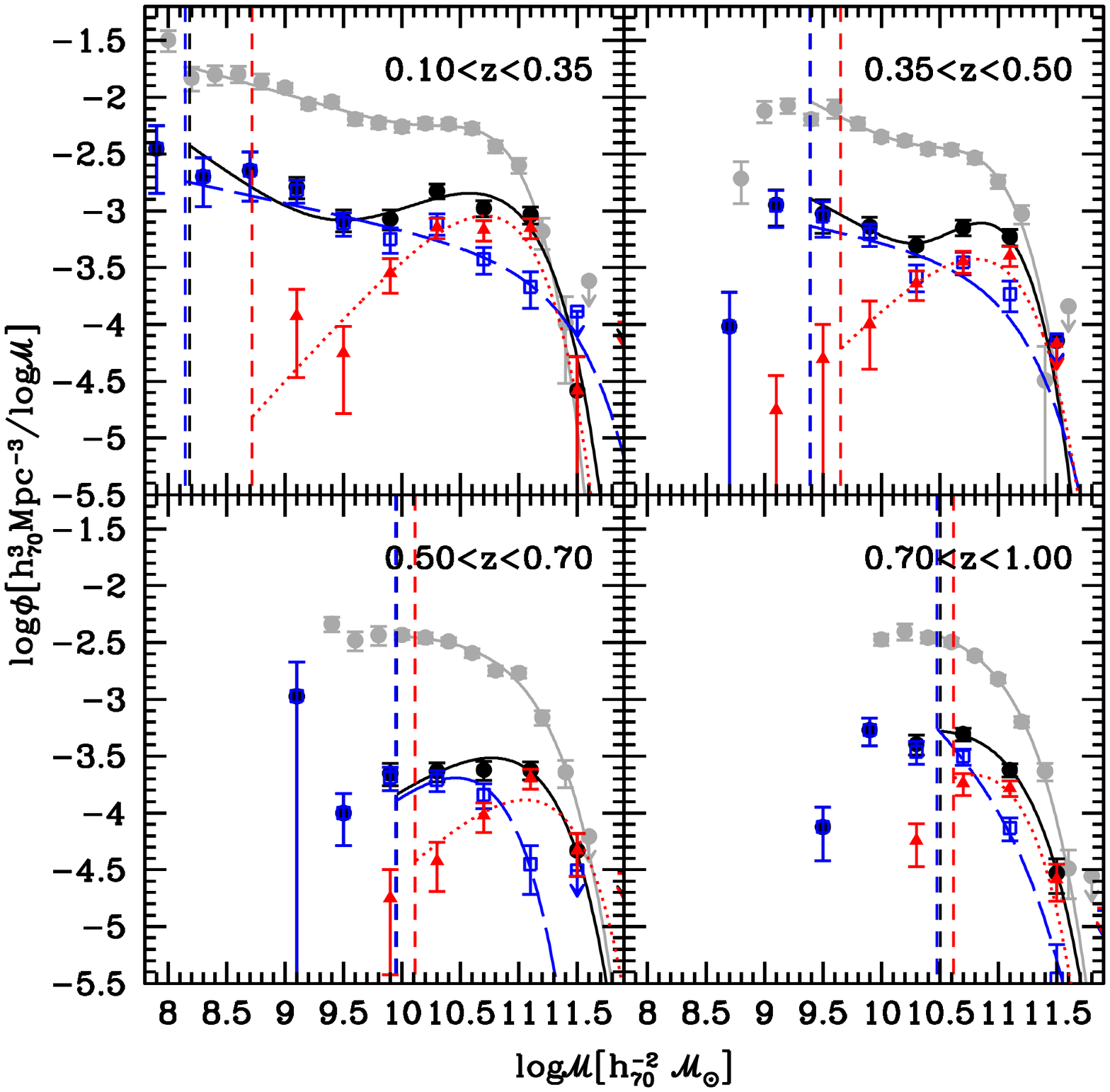}
\includegraphics[width=0.485\linewidth]{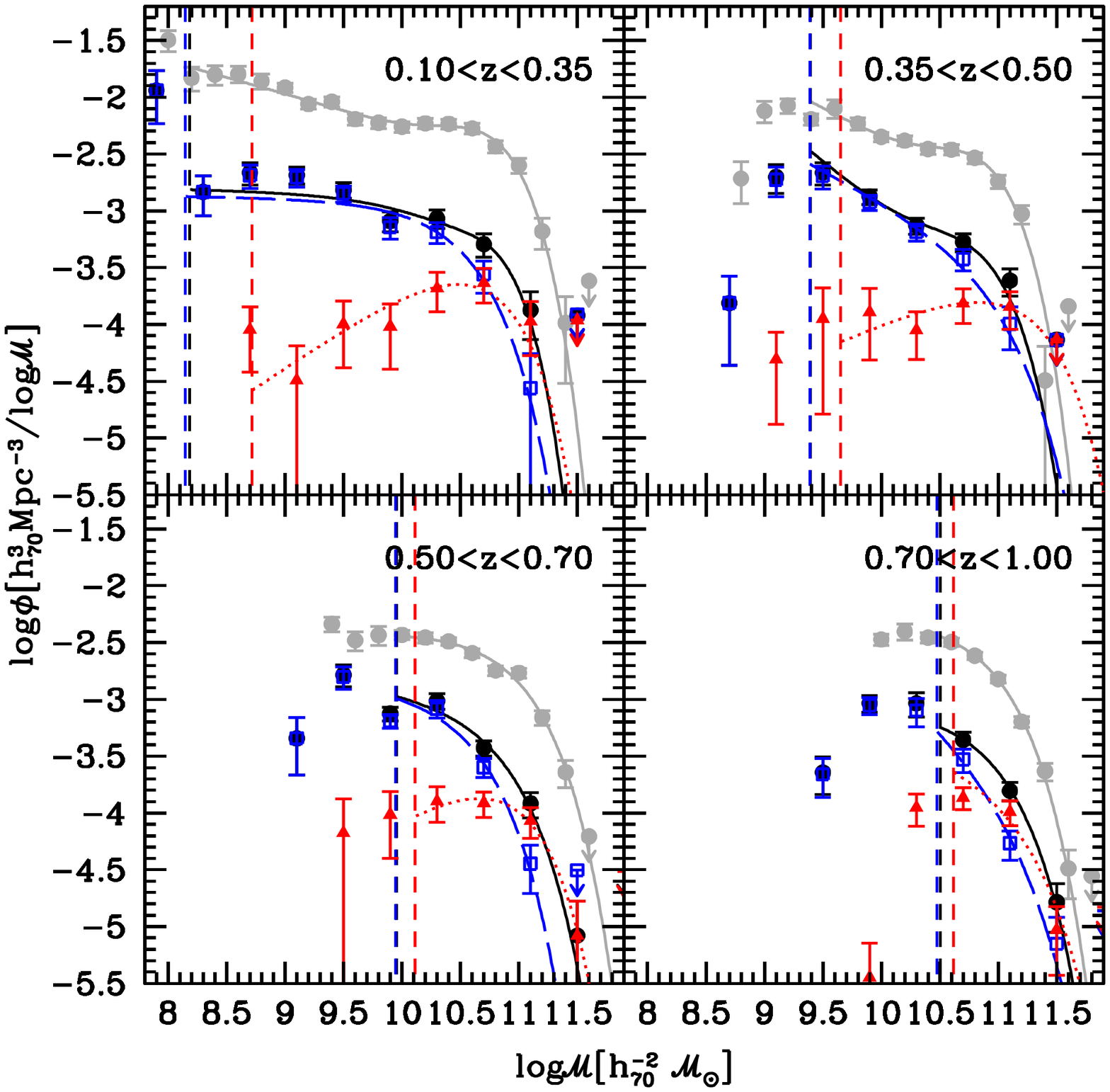}
\caption{The stellar mass function in different environments. The left-hand panels are obtained for the group galaxies residing in the groups with at least 2 corrected $M_B<-20.5-z$ members. The right-hand panels are for the isolated galaxies. The symbols correspond to the $1/V_{max}$ measurements and the continuous lines correspond to the Shechter fits (see text for more details). Each four-panel figure covers $0.1<z<1$, split in four redshift bins indicated on the top. The black, red and blue symbols/curves are results for the populations of all, early type and late type galaxies, respectively. For a comparison, the stellar mass function of the field galaxies is presented with gray symbols/curve. The environmental dependence of the shape of the stellar mass function is explained by the different relative contributions of galaxies of different types in different environments. See \citet{Bolzonella.etal.2009} for the detailed discussion of the role of environment on the stellar mass function.}
\label{fig_gsmf_grisol}
\end{figure*}

For comparison with the literature, the above discussion was based on the rest-frame $B$-band luminosity.  As well as the obvious sensitivity of the $B$-band luminosity to the recent star formation history, there are many reasons to
prefer stellar mass as a yardstick of galaxian properties. In particular, as mentioned in the introduction, the existence of a correlation between
the stellar mass function of galaxies and their environment \citep{Baldry.etal.2006, Bolzonella.etal.2009}, plus the strong correlation between mass and
galaxian properties, means that any sample that contains a 
significant range of masses may show a spurious environmental dependence simply because the different distribution of masses in the
different environments.  This will be true not only for $B$-selected samples, but also $K$-selected samples, and even mass-limited samples (i.e. all galaxies with masses above some threshold).  
The spurious environmental effects produced by this difference in mass functions is shown by the clear flattening of the observed colour-density and morphology-density relations \citep{Cucciati.etal.submitteda, Tasca.etal.2009} when going from narrow $M_B$ luminosity bins to narrow mass bins.

Using the overall zCOSMOS density field \citep{Kovac.etal.2009}, \citet{Bolzonella.etal.2009} have shown that the stellar mass functions of galaxies are different in different environments up to at least $z \sim 0.7$, and probably at higher redshifts. The galaxian stellar mass function in the densest regions, defined by the highest quartile in the distribution of galaxy environments at that redshift bin, is of bimodal shape at least up to $z \sim 0.5$, showing an upturn at low-mass end and with massive galaxies preferentially residing in these densest regions. \citet{Bolzonella.etal.2009} explain the environmental dependence of the shape of the stellar mass function by the different relative contributions of galaxies of different types in different environments. The individual stellar mass function for each of the different types is well represented by the Schechter function in all environments  (see Figure 4 in \citealt{Bolzonella.etal.2009}).  The higher contribution of the red/early objects in the densest environments produce a characteristic ``bump'' at high-mass end in the stellar mass function in the densest environment.

In this paper we are investigating the morphological properties of galaxies in the group environment \citep{Knobel.etal.2009}, corresponding to virialised structures (i.e. single dark matter haloes) - and can be confident from the comparison with the mocks that this is indeed to a large degree the case \citep{Knobel.etal.2009}.  Not surprisingly, galaxies in the identified groups reside also in the denser environments (see Figure 15 in \citealt{Kovac.etal.2009}), and the groups are themselves good tracers of the dense peaks in the density field (see Figure 16 in \citealt{Kovac.etal.2009}). However, there is not a one-to-one correspondence between the group and the continuously defined environment.  

We have therefore recalculated the stellar mass function for the samples of the group and isolated galaxies used in this paper, following exactly the same methodology as described in \citet{Bolzonella.etal.2009}. The stellar mass functions are calculated using the $1/V_{max}$ method, by using the additional weights for each galaxy to correct for the average sampling and redshift success rate of the 10k HCC zCOSMOS sample to the full  40k sample (as the stellar mass function deals with the volume densities and not fractions of objects, similar to the weights used to define the group richness). The errors correspond to Poissonian errors obtained from the $1/V_{max}$ method. The $V_{max}$ refer to the total zCOSMOS volume, and therefore the normalisation of all stellar mass functions, except of the one of the total sample of galaxies, is arbitrary. As for the rest of the analysis in this paper, galaxies are selected to reside in the central zCOSMOS region, defined in Section~\ref{sec_sampledef}.

\begin{table*}
\centering
\begin{tabular}{c|cccc}
\hline
$M_{*,compl} \ limit$ & $M_*>9.88$  & $M_*>10.21$ & $M_*>10.68$& $M_*>10.96$ \\
$z \ range$ & $0.1<z<0.45$ & $0.1<z<0.6$ & $0.1<z<0.8$ & $0.2<z<95$ \\
\hline
$Group (M_B<M_{B,compl})$ & 422 & 390 & 294 & 132 \\
$Group (M_B<20.5-z)$ & 241 & 235 & 218 & 132 \\
$Field$ & 987 & 1045 & 858 & 428 \\
$Isolated$ & 141 & 180 & 109 & 40 \\
\hline
\end{tabular}
\caption{\label{tab_mass}Number of (non-weighted) galaxies in the mass complete samples.}
\end{table*}

The resulting stellar mass functions for the group and isolated sample of galaxies for four redshift bins are shown in the left and right panel in Figure~\ref{fig_gsmf_grisol}, respectively (black symbols and curves). The mass functions are fitted with a double Schechter function (with a unique value of the exponential cut-off) in the two lower redshift bins and with a single Schechter function in the two higher redshift bins. To be able to compare similar group environments at different redshifts, we select only group galaxies residing in the groups with at least 2 corrected $M_B<-20.5-z$ members. The shape of the stellar mass functions in the sample of group and isolated galaxies is reminiscent of the shape of the stellar mass functions in the sample of galaxies in the highest and lowest quartile of overdensities in \citet{Bolzonella.etal.2009}. For the group galaxies, similar results are obtained when using all group members without restriction in the group richness - smoothing slightly the characteristic bimodal shape of the mass function in the dense region in the two lowest redshift bins, as expected when including in the galaxy sample also members of the smaller groups.

We also calculate stellar mass function per morphological type in different environments, using the morphological classification described in Section~\ref{sec_morphdef}. As expected, the high mass end of the stellar mass function in both environments is dominated by the early type galaxies, peaking at roughly $10^{11}$ \Msol. Towards lower masses, the number of early type galaxies decreases and the number of late type galaxies increases. Late type galaxies dominate below the crossing mass (the mass at which the two mass functions cross) and this crossing mass shifts towards lower values at later cosmic epochs. Since $z \sim 0.7$, the crossing mass is clearly lower for the group than the isolated sample of galaxies. We will discuss the crossing mass in more detail in the following Sections.

We therefore find that the stellar mass function exhibits a similar shape dependency on environment, whether environment is defined as a continuous field (density of objects) or a discrete (groups/isolated sample) quantity. Moreover, the reason is the same: the different relative contributions of galaxies of different types across a range of masses (see \citealt{Bolzonella.etal.2009} for more discussion). 

We stress that one consequence of this for any analysis related to the environment, is that whenever there is a sample of galaxies which contains a range of stellar masses, then the relationships between a given galaxy property and environment will be modulated by the relationship between that property and stellar mass through the differing mass distributions of galaxies with environment. 
To isolate the true environmental effect, we must consider galaxies in narrow stellar mass bins.

\begin{figure*}
\centering
\includegraphics[width=0.9\linewidth]{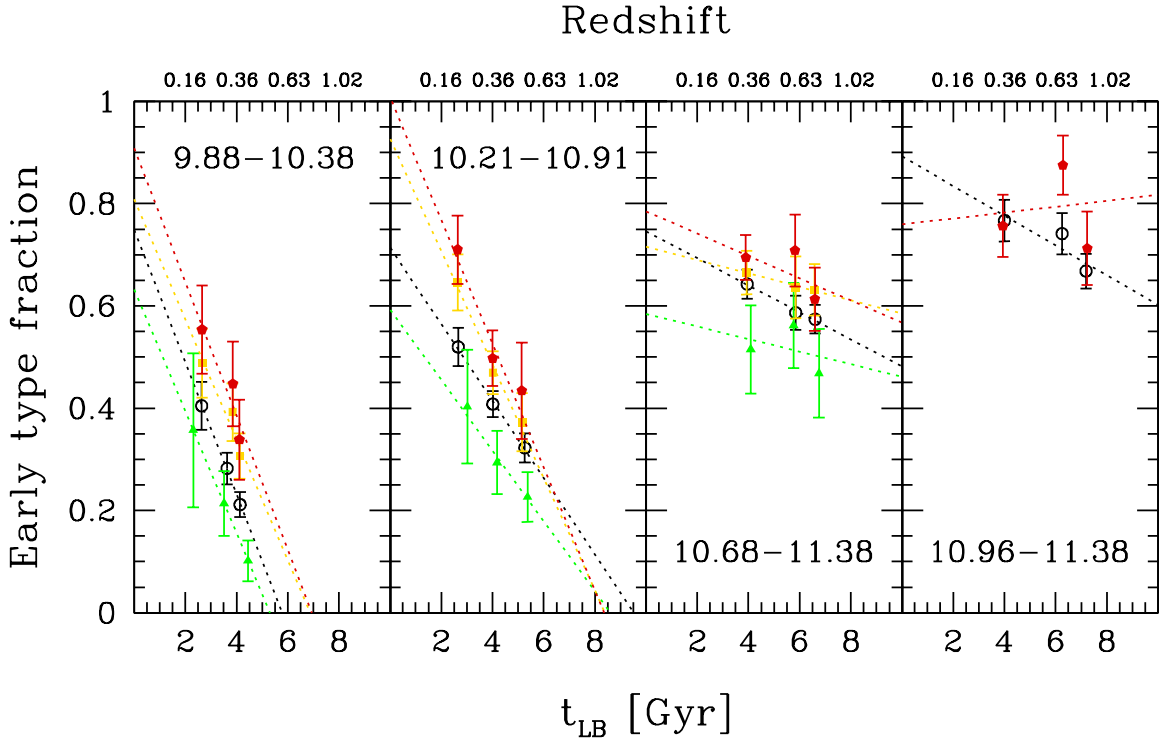}
\caption{Redshift evolution of the fraction of early type galaxies of a given stellar mass in different environments. The stellar masses are from intervals of $\log(M_*/M_{\odot})=0.7$ width, where the interval limits in $\log(M_*/M_{\odot})$ units are given in the individual panels. The lower stellar mass value represents the $85\%$ completeness limit for the early ($1+2.0$) type galaxies in the redshift interval of consideration:  $0.1<z<0.45$, $0.1<z<0.6$, $0.1<z<0.8$ and $0.2<z<0.95$ going from the left to the right, respectively.  The solid yellow squares and solid red pentagons represent group galaxies with an effective richness of at least two in the corresponding luminosity complete sample and in the $M_B<-20.5-z$ sample of galaxies, respectively. The empty circles represent the field galaxies and the green solid triangles represent the isolated galaxies. The symbols are plotted along horizontal axis at the median look-back time/redshift of galaxies in the considered bin. The dotted lines are the linear fits to the $f_{early}-t_{LB}$ relation for the various samples, marked in the same colour as the corresponding early type fractions. For galaxies with $\log(M_*/M_{\odot}) \lesssim 11$, there is a steady increase with cosmic time in the fraction of early type galaxies in all environments. Moreover, the fraction of early type galaxies is highest in the group environment, and lowest for the isolated galaxies at every redshift probed by that particular $\log(M_*/M_{\odot}) \lesssim 11$ stellar mass.}
\label{fig_morphfracmassenv}
\end{figure*}

\subsection{Morphological mix in different environments from the mass selected samples}
\label{sec_masssample}

We  therefore select four  stellar mass-complete  samples  of   galaxies  to
investigate the morphological mix inside and outside groups in zCOSMOS. The selection 
of the complete stellar mass samples  has been described
in \citet{Pozzetti.etal.inprep}, and is optimised to be complete for both early 
and late type galaxies following our definition of morphological types. We use the samples which are 85$\%$ complete for the early (1+2.0) type galaxies. The  samples are defined as  follows: $\log
(M_*/M_{\odot}) > 9.88$ in  $0.1<z<0.45$,  $\log  (M_*/M_{\odot}) > 10.21$  in
$0.1<z<0.6$, $\log  (M_*/M_{\odot}) > 10.68$ in $0.1<z<0.8$ and $\log (M_*/M_{\odot}) > 10.96$ in $0.2<z<0.95$. The numbers of galaxies in the different environments within these mass complete samples are given in Table~\ref{tab_mass}.

The morphological mix of galaxies within
these mass-selected samples in different environments is shown in  Figure~\ref
{fig_morphfracmassenv}. The stellar mass bins have a width of $\Delta \log(M/M_{\odot})=0.7$ (dictated by the number statistics), starting from the lowest stellar mass for which the samples will be complete. This makes our mass bins partially overlapping, and consequently, not independent of each other. 

\begin{figure*}
\centering
\includegraphics[width=0.92\linewidth]{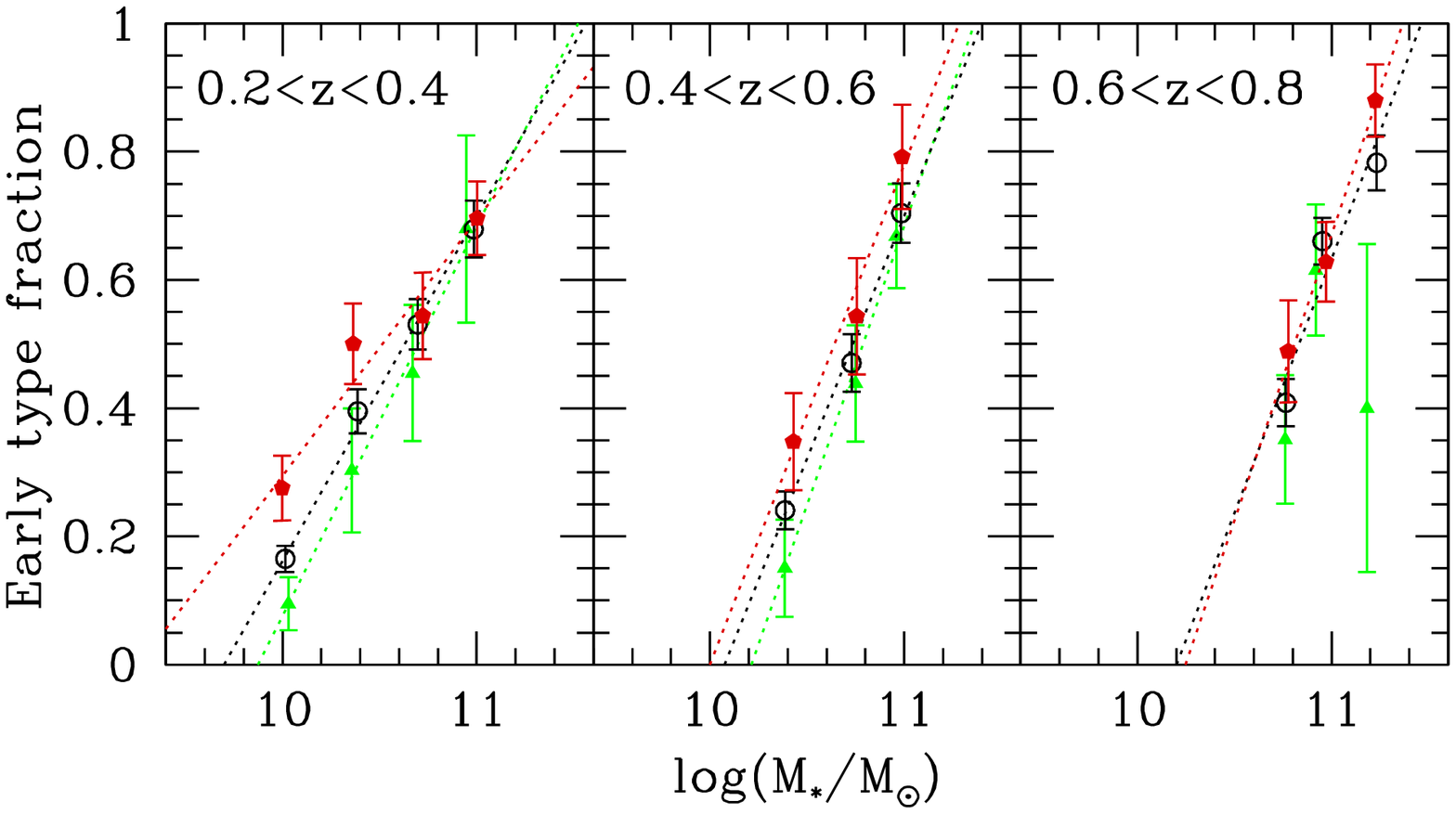}
\caption{Environmental dependence of the fraction of early type galaxies on stellar mass.  The red solid pentagons represent the group galaxies with the corrected effective richness of at least two in $M_B<-20-z$ sample of galaxies, the empty circles represent the field galaxies and the green solid triangles represent the isolated galaxies. In the selected redshift intervals galaxies are extracted from the stellar mass complete samples ($85\%$ completeness limit for the early $1+2.0$ type galaxies). The symbols are plotted along the horizontal axis at the median mass of galaxies in the considered bin. The dotted lines are the linear fits to the $f_{early}-\log(M_*/M_{\odot})$ relations, marked in the same colour as the symbols in the corresponding environments. At every redshift, the fraction of early type galaxies increases strongly with the stellar mass for galaxies residing in both the groups and in the field. The morphological mix that is achieved in the groups at some redshift and at some mass, is achieved at only slightly later times (or equivalently for slightly higher masses) in the field and for the isolated galaxies.} 
\label{fig_morphfracmass_zenv}
\end{figure*}

For the stellar masses below $\sim \log(M/M_{\odot})=11$ we observe a steady increase with cosmic time in the fraction of early type galaxies in all environments. For the highest mass bin, $10.96<\log(M/M_{\odot})<11.66$, the number of galaxies is small, and little can be conclusively said regarding the groups and for the field. However, it seems clear that high fractions $\sim 70-80\%$ of early type $\sim \log(M/M_{\odot}) > 11$ galaxies in both the group and field are present before $z=1$ and that the morphological mix stays 
more or less constant since then.

Progressing to lower masses, we find that, at a given stellar mass, the fraction of early type galaxies is consistently
highest in the group environment, and lowest for the isolated galaxies, at every redshift that we can examine for that particular stellar mass. Fitting a linear relation between the early type fraction and look-back time, the linear fits for galaxies with $\log(M/M_{\odot}) \lesssim 11$ suggest that the fractions of early type galaxies were more or less the same in all types of environment at higher redshifts and then, with cosmic time, diverged with progressive morphological transformation from late to early type galaxies happening at different rates in different environments at a given mass.  
As above, there is an indication that the increase in the early type fraction over cosmic time is more rapid in the groups with higher richness than in the poorer groups (e.g. difference between red pentagons and yellow squares and the corresponding linear fits in Figure~\ref{fig_morphfracmassenv}).

Looking at the same data from another perspective, we can calculate the fractions of early type galaxies as a function of stellar mass in the narrow redshift bins: $0.2<z<0.4$, $0.4<z<0.6$ and $0.6<z<0.8$. These are shown in Figure~\ref{fig_morphfracmass_zenv}, using only groups with a corrected richness of at least two in $M_B<-20-z$ galaxies, as we extend this analysis only to $z=0.8$.  At every redshift, the fraction of early type galaxies increases strongly with the stellar mass for galaxies residing in both the groups and in the field. The isolated galaxies follow the same trend in the lowest redshift bins up to $z=0.6$, while in the highest redshift bin they become very rare objects and the obtained statistics is much less reliable.  This figure also illustrates the convergence of the morphological mix at a given (relatively high) mass at redshifts approaching $z \sim 1$.

These results are in good agreement with \citet{Tasca.etal.2009}, who showed the weakening of the overall morphology-density relation at high stellar mass and/or high redshifts in similar mass complete samples in the zCOSMOS. However, the increase of the early type fraction with stellar mass in $0.2<z<0.8$ (Figure~\ref{fig_morphfracmass_zenv}) contrasts results obtained by \citet{Holden.etal.2007}, who do not detect any clear trend in the early type fraction with stellar mass for the cluster galaxies more massive than $\log(M/M_{\odot})=10.52$. We have taken into account the difference in the IMF used by \citet{Holden.etal.2007} and here, assuming that the stellar masses calculated with the Chabrier IMF are larger by 0.08 than the ``diet Salpeter'' IMF used by \citet{Holden.etal.2007} (see e.g. \citealt{Bolzonella.etal.2009} on the differences between the stellar masses with the different IMFs). \citet{Holden.etal.2007} results are obtained for five massive X-ray clusters in $0.023<z<0.83$ spanning a range in velocity dispersion $865 < \sigma < 1156$ \kms, or with 56 to 109 members above the quoted mass limit. The measured fractions of early type galaxies are between 71$\%$ and 100$\%$, comparable to the fractions which we measure only for the most massive $> \log(M/M_{\odot}) \sim 11$ galaxies. This indicates that clusters are either more efficient in transforming late types to early types or that the bulk of  morphological transformation of galaxies more massive than $\log(M/M_{\odot})=10.52$ has happened earlier in the clusters than in any other environment, and specifically the poorer group environments examined here.

\begin{figure}
\centering
\includegraphics[width=0.95\linewidth]{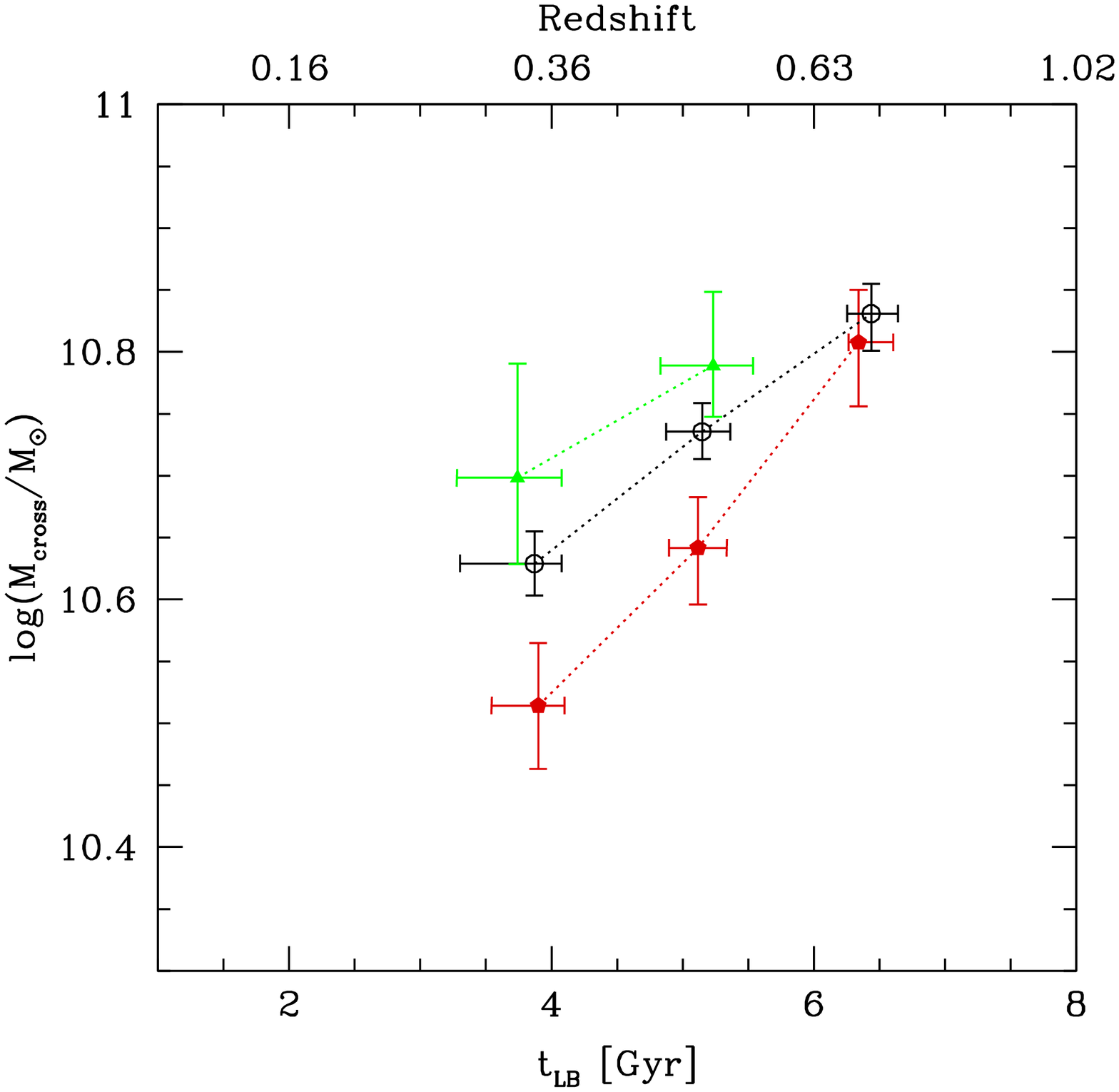}
\caption{Morphological crossing mass of galaxies in different environments. The red solid pentagons represent the group galaxies, the empty circles represent the field galaxies and  the green solid triangles represent the isolated galaxies. The crossing mass is the mass at which 50$\%$ of the selected galaxy population is of early (or equivalently of late) type. The stellar masses are estimated from the linear fit to the measured points in Figure~\ref{fig_morphfracmass_zenv} in the individual redshift intervals.  The vertical error bars correspond to the rms values in the crossing mass obtained by bootstrapping of the galaxy samples and repeating the fitting, using the previously estimated errors on the real fractions. The horizontal error bars correspond to the upper and lower quartiles in the look-back time  distribution of galaxies in the considered redshift bins, plotted at the median look-back time/redshift in that bin. The morphological crossing mass decreases with decreasing redshift, indicating that the morphological transformation from late to early types starts in galaxies with higher stellar masses and shifts to galaxies with lower stellar masses over cosmic time. Moving toward the lower redshifts, the morphological crossing mass decreases more rapidly for galaxies residing in groups than for the field and isolated galaxies.}
\label{fig_morphfracmass}
\end{figure}

Figure~\ref{fig_morphfracmassenv} and Figure~\ref{fig_morphfracmass_zenv} also show one way of looking at the relative importance of mass and environment in
driving morphological transformation.  It can be seen that the morphological mix that is achieved in the groups at some redshift and at some mass, is achieved at only slightly later times (or equivalently for slightly higher masses) in the field (all galaxies) and in the isolated galaxies. At $z \sim 0.5$ and $\log(M/M_{\odot}) \sim 10.2$, the ``effect'' of group environment is equivalent to only about 0.2 dex in stellar mass or to about 2 Gyr in time.

Fitting the linear relation to the observed $f_{early} - \log(M/M_{\odot})$ trends at a given redshift bin, we can calculate the stellar mass at which $50\%$ of galaxies in a given sample is of early type at that redshift. We call this quantity the morphological crossing mass since it is also obviously the mass at which the individual mass functions of the two populations cross.  The change with redshift of the morphological crossing mass in the group and other environments is shown in Figure~\ref{fig_morphfracmass}.  The morphological crossing mass decreases with decreasing redshift, indicating that the morphological transformation from late to early types starts in galaxies with higher stellar masses and shifts to galaxies with lower stellar masses over cosmic time. This is another manifestation of the so-called downsizing scenario in the evolution of galaxies \citep[e.g.][]{Cowie.etal.1996}. At $z \sim 0.7$ or a little above, $50\%$ of galaxies with masses $\log(M/M_{\odot}) \approx 10.85$ are already of the early type, independent of their environment. Moving toward the lower redshifts, the morphological crossing mass decreases more rapidly for galaxies residing in groups than for the field and isolated galaxies reflecting the emergence of morphological segregation.

\section{Discussion} 
\label{sec_discussion}

\subsection{The relative role of the stellar mass and environment in determining galaxian properties}

The above results have painted a consistent picture of a morphological segregation emerging between the typical group and field environments (i.e. all galaxies) emerging over the last several billion years, since $z \sim 1$, and of being more prominent in galaxies of lower luminosity and/or stellar mass. Corresponding trends in the zCOSMOS 10k sample relative to the larger scale density field were shown in \citet{Tasca.etal.2009}.   

On the one hand, it is clear that this emerging environmental dependence is superposed on top of a global and strong mass-driven evolution that is seen in the whole population. There is no doubt that the stellar, or possibly total baryonic, mass (``nature'') is directly related to the processes responsible for this morphological transformation. The offset between the observed fractions of different morphological types of galaxies in different environments since $z \sim 1$, i.e. higher fractions of early type galaxies for the group than the typical field population, which are again higher than the fractions for the isolated galaxies (Figure~\ref{fig_morphfracmassenv}), is however tuned by the environment. As noted above, at a given mass, this environmental tuning corresponds to about 2 Gyrs in time: a certain morphological mix, a product of the efficient morphological transformation, will be achieved first in the group, then in the field and at the end in the isolated environments for galaxies of a given mass. 

However, it is not immediately clear if this is because of an offset in ``starting time'', as the more massive galaxies are expected to be formed first in the more massive structures (again ``nature'', e.g. \citealt{Kauffmann.1995, Benson.etal.2001, Heavens.etal.2004}), or because of a difference in the rate of evolution in different environments (``nurture''), caused by some of the dense environment related processes (e.g. ram pressure stripping, harrasment etc, see Introduction).

Moreover, we do clearly detect an apparent increase in the environmental differences, at least for the lower mass galaxies which are transforming at the lower redshifts. This produces the clear divergence of the morphological crossing masses from $z \sim 0.7$ to $z \sim 0.3$ in different environments, suggesting that the morphological transformation rate is rather faster in the groups than outside. However, interpretation of Figure~\ref{fig_morphfracmass} in terms of the physically relevant processes is not trivial, since it also reflects both the shape of the mass functions and their relative amplitudes in different environments. 

We therefore investigate below more directly the rates of the morphological transformation. It is also of interest to 
compare the transformation rates for other relevant transformations, and specifically those 
related to the star-formation history, i.e. the
``quenching'' transformation from blue to red, or from active star-formation to passive evolution.  Of course, all of these processes may be reversible, and so we are really looking at the ``net'' transformation rate.  Comparison of these transformation rates may provide clues as to the physical linkage between morphological transformation and star-formation quenching.

\subsection{Timescales for the type-transformations of galaxies}

One interesting question is whether it is possible to determine whether one of morphological transformation and
star-formation quenching preceded the other.  Based on the field stellar mass functions of the zCOSMOS galaxies of different types, \citet{Pozzetti.etal.inprep} find that the morphological crossing mass $M_{sm, cross}$ has higher values than the equivalently defined crossing mass for the populations of photometrically red and blue, and also active and passive galaxies. From the differences in the number density of photometrically and morphologically early type galaxies, the inferred delay time between the colour and morphological transformation is about 1-2 Gyr. A similar timescale is derived for the transformation to red colours after switching of the star-formation. \citet{Bundy.etal.2006} find the qualitatively same behaviour for the crossing mass of the DEEP2 field galaxies when population of galaxies is split according to the morphology and colour, extending the colour/morphology dichotomy of crossing mass to redshift 1.4.  

Similar results have been obtained for the population of cluster galaxies. \citet{Wolf.etal.2009} study the population of galaxies in A901/2 cluster complex at $z\sim 0.17$, finding the majority of high mass galaxies (above $10^{11}$ \Msol) to be red and spheroidal in all environments (outskirt, centre and outside of cluster). Based on the large number of cluster spirals in the mass range of $10^{10}-10^{11}$ \Msol, \citet{Wolf.etal.2009} conclude that quenching of star-formation in these galaxies is a slow process, where the timescale of morphological evolution must be longer than that of the spectral evolution. However, at even lower masses, not accessible by our study, star-formation rate decline and morphological changes appear to be more synchronised in A901/2. Using the sample of 24 clusters from the EDiSCs \citep{White.etal.2005}, \citet{Sanchez-Blazquez.etal.2009} find that the fraction of early type red sequence galaxies decreases by $\sim 20\%$ from $z=0.75$ to $z=0.45$, what can be expected if the cluster spiral galaxies first get redder, stopping their star formation, before they become of early type (note that \citealt{Sanchez-Blazquez.etal.2009} have done the analysis using the luminosity complete samples).   

Using the marked correlation statistics on the SDSS data, \citet{Skibba.etal.2008} conclude that $z \sim 0$ red spirals are often satellites in the outskirts of groups and clusters, while red early type galaxies are preferentially located in the centres of these structures. \citet{Skibba.etal.2008} suggest that the timescale of the morphological transformation of these red spirals is longer than that of the quenching of their star formation. The zCOSMOS extends a high redshift (up to $z \sim 1$) comparison between the colour and morphological properties of galaxies to the group environment.

\begin{figure}\centering
\includegraphics[width=0.9\linewidth]{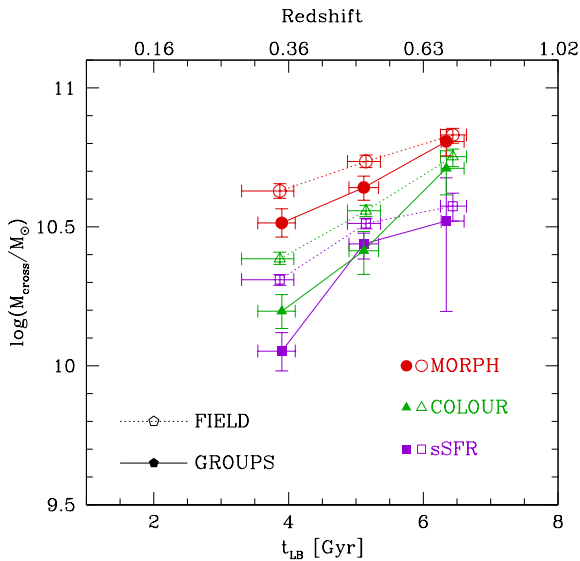}
\caption{Evolution of the class-dependent crossing mass in the group and field environments. The filled symbols are for the group and the empty symbols are for the field samples. The red circles, green squares and violet squares correspond to the morphological, colour and star-formation activity crossing mass, respectively. Error bars for each value have the same meaning as described in Figure~\ref{fig_morphfracmass}. The effective transformation of galaxies from active to passive, from blue to red and from late to early moves progressively towards lower mass galaxies as cosmic time passes or as the environment is denser. The crossing masses in morphology are higher, and change less rapidly, than those for colour and/or star-formation activity.}
\label{fig_crossingmass}
\end{figure}

To enable a closer comparison between the colour and morphological properties of galaxies in the group environment, we calculate the colour crossing mass using the same procedure and the same galaxy samples as for calculation of the morphological crossing mass, described in Section~\ref{sec_masssample} (see also \citealt{Iovino.etal.submitted} who calculate a similar quantity, $t_{50-50}$, the cosmic time at which the fraction of red/blue galaxies is 50$\%$). In addition, we also calculate the crossing mass for star-formation activity by partitioning galaxies into active and passive classes using the specific star-formation rate (sSFR) from the SED fitting (see discussion in \citealt{Pozzetti.etal.inprep} on the agreement between the star-formation rates derived from these SEDs and those from measurement of emission lines).  In this work, we consider galaxies with $U-B>1.1$ to be red, and galaxies with $\log(sSFR/yr) < -10.5$ to be passive. These crossing masses for the group and field samples are shown in Figure~\ref{fig_crossingmass}.

The crossing mass for each of the samples shows similar behaviour as we have shown already for the morphological crossing mass. It shifts to lower masses with cosmic time in both the field and group environment, and since $z \sim 0.6$, this evolution is clearly environment dependent. The activity crossing mass is systematically lower than the colour crossing mass, which is systematically lower than the morphological crossing mass. This is observed both in the field (in agreement with \citealt{Pozzetti.etal.inprep}) and in the group environment. Extrapolating the plot, at a given mass 50$\%$ of population will turn to be passive at earlier cosmic time than the time at which  50$\%$ of population will become red.  50$\%$ of population will become of early morphological type even later. Moreover, this will generally happen earlier for a group galaxy than for a field galaxy.  A similar sequence of the colour and morphological crossing mass in the zCOSMOS has been presented in \citet{Bolzonella.etal.2009} for the environments of extreme densities, when quantifying environment using the local overdensity of neighbouring galaxies \citep{Kovac.etal.2009}. \citet{Bundy.etal.2006} find a tentative evidence for the rise of the quiescent population in dense, continuously defined environments.

While the exact values of the crossing mass are strongly dependent on how galaxies were divided into types, there is a clear and progressive (with epoch) environmental dependence in these values. The effective transformation of galaxies from active to passive, from blue to red and from late to early moves progressively towards lower mass galaxies as cosmic time passes or as the environment is denser. Derived values of the various crossing masses are consistent with a scenario in which, for the majority of galaxies, the timescale for the morphological transformation is longer than the timescale for the colour transformation, both in the field and group environment, assuming that all these processes started to act approximately at the same time in a given environment.

Unfortunately, it is hard to use arguments such as the location and rate of change of the above-defined crossing mass to infer quantitative information on the transformation rates, since these may be sensitive to the choice
of the location of the dividing line between the two properties and also on the shape and relative normalisation of the mass functions of different components of the population. The dividing line may be clear for colour (e.g. the so-called ``green valley'') but 
more arbitrary in the case of morphological classification. The relative amplitude of the mass functions of early or late type 
galaxies, and of blue or red galaxies, or of active or passive galaxies, will also change the location and movement of $M_{cross}$.  
Use of the redshift/epoch axis directly to determine
transformation rates, i.e. of the flux of galaxies crossing through a chosen dividing line, is likely to be less sensitive to 
the precise location of that divide and to allow a more direct study of these questions.

\subsection{Role of group environment in the general galaxy type-transformation rate}

In the following, we explore more directly the ``transformation rate'' of late type galaxies into early type galaxies for different masses and different environments in order to try to better understand the physical causes.  It should be noted in the following discussion that this is really a ``net'' transformation rate from late to early, since it is in principle possible for early type galaxies to reacquire a gaseous disk and become of later morphological type, although whether this would be sufficient to cross our defined morphological divide between ZEST classes 2.0 and 2.1 is unclear.

Assuming from now on that galaxy transformations are dominated by one-way transformation late to early (and blue/active to red/passive), we proceed to
define a normalised transformation rate $\eta$ in which the rate of change in the number of ``progenitor'' galaxies (either late type, blue or active) is normalised
by the number of these progenitors:

\be
\eta = - \frac{1}{N_C} \frac{dN_C}{dt} = - \frac{d \ln N_C}{dt}
\ee

\noindent
where $N_C$ is a measure of the number of progenitors of a given class $C$, i.e. the mean comoving density averaged over suitably large volumes of the Universe. 

The inverse of $\eta$ gives the transformation timescale $t_{\eta}$ for a typical ``progenitor galaxy'' to be transformed.  The transformation rate $\eta$ might be expected to be a function of mass, environment and probably epoch.

Unfortunately, an accurate
determination of $N_C(z)$ and $dN_C/dt$ for a particular class $C$ of galaxies is observationally challenging.  It requires an accurate measurement of galaxian
properties, e.g. the stellar mass, over a range of redshifts, careful treatment of large scale density variations within surveys due to large scale structure, and,
ideally, knowledge of the changes in the number of galaxies in a particular class due to, for instance, mass growth through star-formation and/or merging.  In the case of galaxies defined by environment, i.e. those in groups, there will also be the change in the population due to the hierarchical growth of the groups through the accretion of galaxies onto the virialised structures.  

In cases where it can be assumed that the total number of galaxies in a given bin (defined by mass and environment) is more or less constant, then some of these difficulties can be alleviated by considering the fraction of galaxies, rather than their absolute number, i.e. by dividing by a fixed total number of galaxies $N$.  In this case:

\be
\eta= - \frac{1}{f_C} \frac{df_C}{dt} = - \frac{d \ln f_C}{dt}
\label{eq_eta}
\ee

\noindent
If the evolution of $\ln {f_C}$ is linear with cosmic time, we simply obtain $\eta$ as the slope to this relation for any sample defined by mass and environment. With the limited baseline in redshift and limited number statistics, we will henceforth consider 
the rate $\eta$ as single time-averaged rate over the accessible range of epochs, from a linear fit to $\ln {f_C}$.

For group galaxies, we know that the total number of galaxies is certainly not conserved, because galaxies will be accreting onto the group from the surrounding regions. It is clear that this migration into the groups is likely to further amplify any environmental difference in transformation rates, because the new arrivals will have to ``catch up'' with the existing group members which already have a more evolved state.

In the following, we derive a simple first order correction accounting for the galaxies infalling to the groups by assuming (a) that the total number of galaxies in the sample (at a given mass)
is conserved, i.e. we do not consider galaxy mergers and neglect the effects of star-formation, and (b) that the infalling galaxies can be considered to be a
representative subset of the non-group population. This latter condition may not be true, and this infall correction is therefore likely to represent an upper bound to the actual transformation rate.

In some small increment of time $dt$, the number of class-transformations $dN$ in groups will be given by

\be dN = -N_G df_{CG} - (f_{CG} - f_{CN}) dN_G  \ee

\noindent
where $N_G$ is a number of the group galaxies, $f_{CG}$ and $f_{CN}$ are fractions of the $C$-class galaxies in the groups and non-groups, respectively, and where the class $C$ is a ``progenitor'' type of galaxy (i.e. assumed to be late morphological type or active/blue type). The equation above can be further written as

\be dN = -d(N_G f_{CG}) + f_{CN} dN_G  ~.\ee

\noindent
By normalising the number of transformations with $N_G f_{CG}$, which is the ``pool'' of potential objects in the group to be transformed, the rate $\eta_{IN,CORR}$ including the effects of infall can be written as:

\be
\eta_{IN,CORR} = - \frac{1}{f_{CG}} \frac{df_{CG}}{dt} + \left(\frac{f_{CN}}{f_{CG}} - 1 \right) \frac{1}{f_{G}} \frac{df_G}{dt} ~. 
\label{eq_etacorr}
\ee

\noindent
In the case of no infall, where $f_G$ is constant over time, or in the case where the group and non-group populations have the same 
properties at a given epoch, it can be seen that the rate $\eta$ will be given just with the first term on the left hand side, which is the same as Equation~\ref{eq_eta} above.   

What about progenitor bias?  By considering two simple complementary categories, group and non-group and the change in the fraction of galaxies associated with each, we automatically deal with both additional membership of pre-existing groups and the emergence of new groups where previously none existed (above the fixed threshold in richness). We should however recognise that the rates so derived will be averages across the entire, evolving, group population and not, necessarily, representative of what is 
happening within a single group. Because of the attraction of using two complementary populations, group and non-group, we no longer consider either the isolated galaxies nor the field (all galaxies) populations in this analysis.

\begin{figure}
\centering
\includegraphics[width=0.95\linewidth]{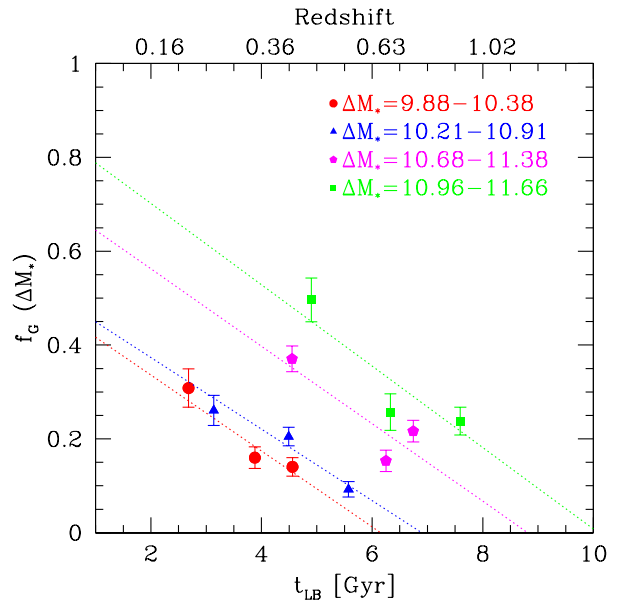}
\caption{Evolution of the fraction of galaxies in our group sample. Different symbols correspond to the different $\Delta \log (M_*/M_{\odot})$ mass bins, indicated in the right corner of the plot. The dotted lines are the linear fits to the time evolution of the fractions of group galaxies, extrapolated to all epochs. At a given stellar mass, the fraction of galaxies residing in the group environment increases with cosmic time, as individual groups grow through infall.} 
\label{fig_infall}
\end{figure}

It is important to note that, based on extensive comparisons with the mock catalogues, neither the completeness nor purity of the group catalogue are believed to depend significantly on redshift \citep[][see Figure 9]{Knobel.etal.2009}.  The same is also true 
of the individual group members. This allows us to use
the observed fraction of galaxies in a volume-limited sample of groups as a good estimate of the infall rate from the 
non-group into the group population, and the observed morphological mix of galaxies in and outside of those groups as representative of both populations. The redshift evolution of the fraction of galaxies residing in the groups with at least two corrected members above $M_B<-20.5-z$ is shown in Figure~\ref{fig_infall}. At a given stellar mass, the fraction of galaxies residing in the group environment increases markedly with cosmic time, building up these structures. At a given time, a larger fraction of the more massive galaxies are already present in the groups reflecting the different mass functions shown above (see Section~\ref{sec_gsmf} and Figure~\ref{fig_gsmf_grisol}).

\begin{figure*}
\centering
\includegraphics[width=0.475\linewidth]{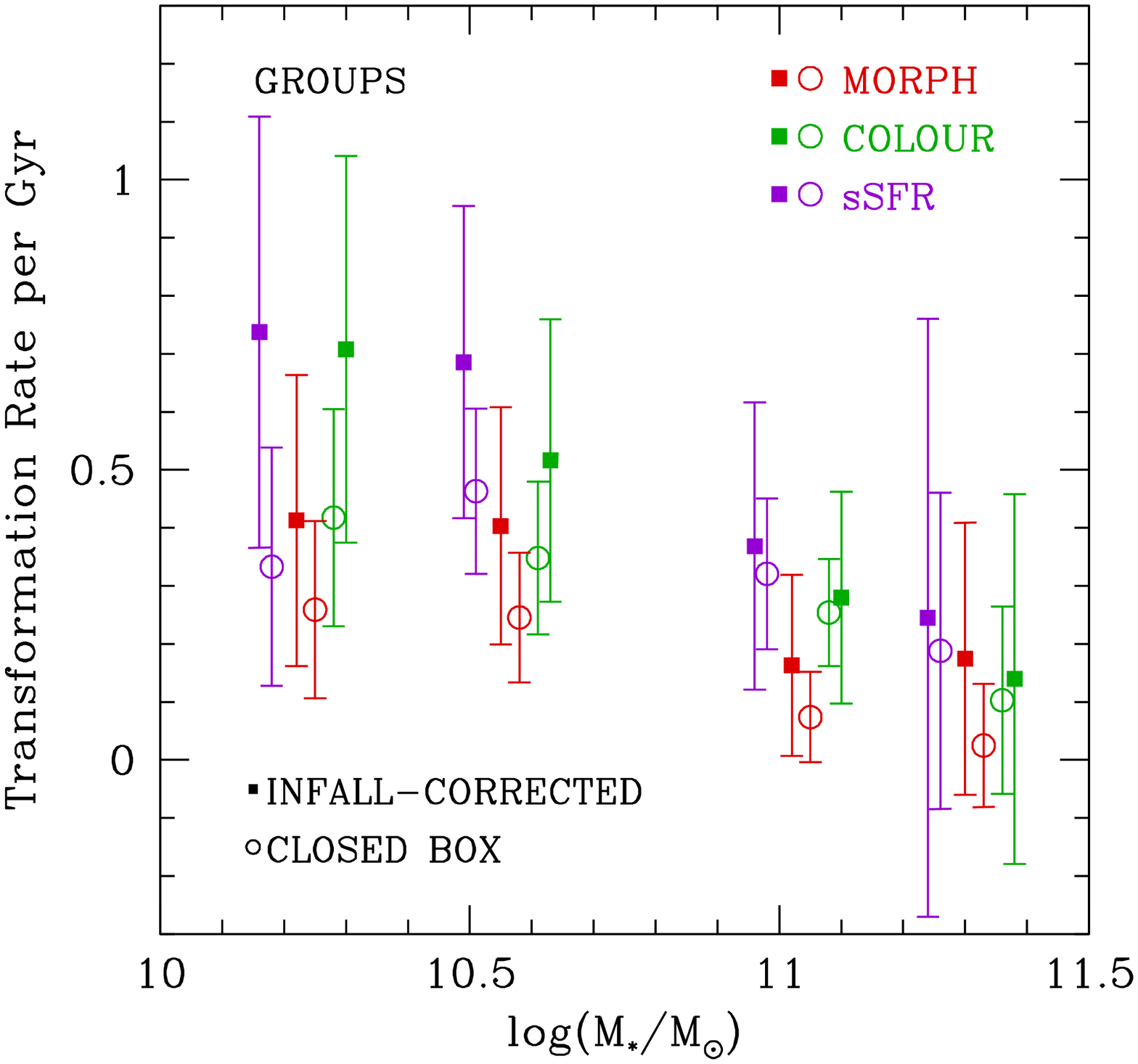}
\includegraphics[width=0.475\linewidth]{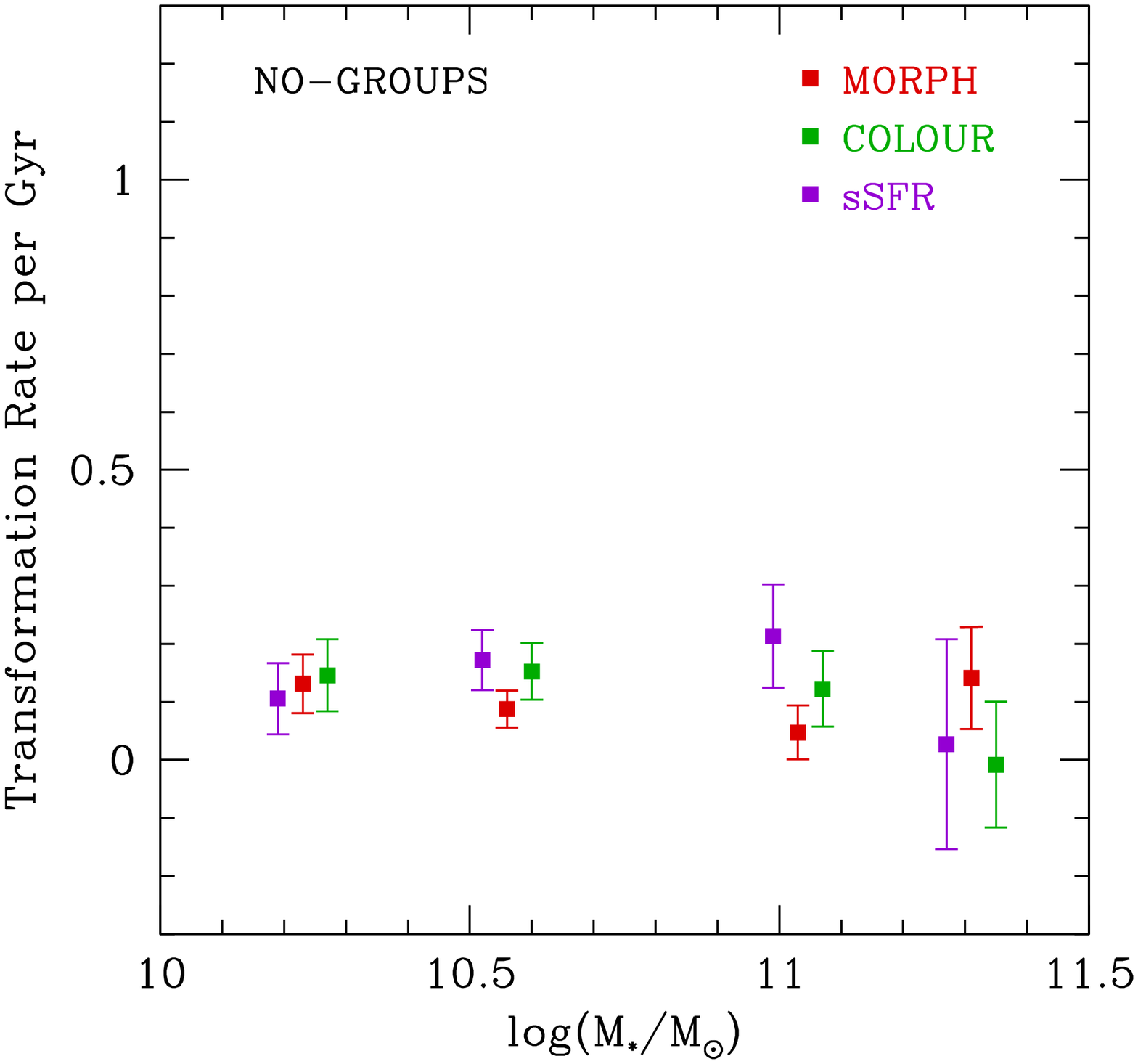}
\caption{Derived transformation rates for morphology, colour and star-formation activity in the group environment (left) and in the non-group environment (right). The solid symbols correspond to the final rates, which are corrected in the case of the group galaxies, for the infall of new members onto existing groups and the appearance of new groups above the selection threshold (see Equation~\ref{eq_etacorr}). It should be noted that this is ``worst-case'' correction which assumes that the new group members are representative of the previous non-group population.  Using this same assumption, the rates in the non-group population need not be corrected for the loss of these galaxies (see Equation~\ref{eq_eta}). The open symbols in the left hand plot also correspond to the rates in the groups with no infall correction applied (Equation~\ref{eq_eta}) for the group galaxies. The morphological, colour and star-formation activity rates are shown in red, green and violet, respectively. The transformation rates are consistently higher in the group environment than outside of it. The rate of transformation is also consistently higher for the transformations involving star-formation than those involving morphological transformation. Finally the rate is higher for galaxies around $10^{10.5}$ \Msol\ than for higher mass ones. The rates must also fall again at lower masses, which cannot yet be studied.}
\label{fig_rates}
\end{figure*}

The derived rates of the morphological transformation $\eta_{L->E}$ in the group environment are shown in the left-hand panel in Figure~\ref{fig_rates}. We obtain the infall corrected rates taking for the term $\frac{f_{CN}}{f_{CG}}$ the average of the measured fractions at different times at a given mass, discussed in Subsection~\ref{sec_masssample} (see Figure~\ref{fig_morphfracmassenv}). The errors are obtained by the error progression of Equation~\ref{eq_eta} and Equation~\ref{eq_etacorr}, using the individual bootstrap errors on the fractions.  The red filled squares represent the rates corrected for infall using Equation~\ref{eq_etacorr}, while the red open circles 
represent those computed directly from
Equation~\ref{eq_eta}, which are lower, as expected. The equivalently derived rates for the non-group galaxies, which are galaxies not residing in the groups with the corrected richness of at least two $M_B<-20.5-z$ members, are shown in the right-hand panel in Figure~\ref{fig_rates} (red filled squares). We use the simple Equation~\ref{eq_eta} above, 
since the infall is a one-way depletion of the non-group population, again assuming that the infalling migrants are representative of the 
population as a whole.

In a similar manner, we derive the rates of the colour transformation from blue to red galaxies $\eta_{B->R}$ and from active to passive galaxies $\eta_{A->P}$ using the same group and non-group samples of galaxies as for measurements of $\eta_{L->E}$.  Galaxies are divided in different classes as in the previous subsection. The corresponding rates are overplotted in Figure~\ref{fig_rates} (using the green symbols for $\eta_{B->R}$ and violet symbols for $\eta_{A->P}$). Generally speaking, the rates obtained for the colour and activity type-transformation also show similar dependence on the mass and environment as the rates of the morphological transformation.

Comparison of the two panels of Figure~\ref{fig_rates} shows three interesting facts:

(a) The rate of the class-transformation is consistently higher in the group environment than outside of it.
(b) The rate of transformation is consistently higher for the transformations involving star-formation than for those involving morphological transformation.
(c) The rate is higher for lower mass galaxies than for higher mass ones. Indeed the rates, and the differences between the rates inside and outside of groups, and
between morphology and star-formation, are all small and consistent with zero for the most massive galaxies above $\log(M_*/M_{\odot}) \sim 11.2$ \Msol. It should be noted that these rates must fall at still lower masses, beyond the range explored by zCOSMOS.

For typical galaxies with $\log(M_*/M_{\odot}) \sim 10.5$, we conclude that the transformation rates are 3-4 times higher in the groups than outside, 
and are typically 50$\%$ higher for colour transformations than for morphological ones.  The implied transformation timescales $\eta^{-1}$, are of order 1.6 Gyr (colour transformation) or 2.5 Gyr (morphology transformation) in the groups, and about 6 Gyr and 10 Gyr, respectively, outside of groups. 

Needless to say, the exact values of the derived rates and corresponding timescales should be taken as indicative only. Not least, the quoted errors include only the purely statistical uncertainty. There are various other issues which one needs to be aware of. Although we have tried to deal with variations in $N(z)$ due to the large scale structure by considering fractions, the temporal gradients may be affected by the choice of redshift bins, since this will be affected by large scale structure. At low $z$, the volume used for the analysis is the smallest, and the estimated fractions could be less representative of the universal values than the fractions estimated at higher redshifts, modifying the the true slopes in the overall fractional evolution. Moreover, we have assumed constant rates (i.e a linear evolution in the logarithmic fraction), and have furthermore estimated these at different epochs for the different mass ranges. We believe that the relative values should be more robust than the absolute ones.

It is fairly clear, e.g. comparing with Figures~\ref{fig_gsmf_grisol} and ~\ref{fig_morphfracmassenv} that the difference in the transformation rates computed here stem as much from the different fractions in the different different environments
(in the denominator) as from differences in the rate of change of the fractions (in the numerator), which are often rather similar. It is also clear that this difference in the fractions between the different environments amplifies the difference in implied transformational rates in the groups through the infall correction, increasing the difference by about $\sim 50\%$. However, it is clear that this is not the sole origin of the rate difference between the different environments.

It should be noted that there are at least two effects that would act to reduce the observed rates, and therefore lead us to underestimate the effect of the group. First, as noted earlier in the paper, 
the non-group galaxies as we define them here, will be somewhat ``polluted'' by galaxies in small groups that are not yet detectable in the relatively sparse sampled zCOSMOS 10k sample.  This will tend to wash out the environmental differences. This contamination will arise from the non-perfect isolation of the
group sample in terms of the imperfect purity and completeness of the original group catalogue.   
Second, we have implicitly assumed that the number of galaxies is conserved.  Certainly in the case of the morphological transformation, galaxy 
merging is a possible physical mechanism. It could well be that the merger of two late type galaxies would produce only one early-type, and the implied change in the late-type fraction would then be smaller than if two early type galaxies had been produced, leading to an underestimate in the transformation rate as defined in Equation~\ref{eq_etacorr}.   Against these, it should be remembered that the ``infall correction'', which also accounts for the appearance of new groups with time, will be a worst case, because it assumes the new group galaxies are drawn randomly from the earlier non-group population.

The derived transformation rates are of course closely linked to the mass functions given above (Figure~\ref{fig_gsmf_grisol}). In particular, given the existence of the transition or crossing mass (where the population is split equally between the two types), and the evidence for change of this with epoch, we would expect to see the highest transformation rates for galaxies around this same mass $\log(M_*/M_{\odot}) \sim 10.5 \pm 0.5$.  At higher masses the transformation rates are lower because the population has saturated with most transformations completed, more than compensating for the small number of potential progenitor galaxies.  At lower stellar masses, $\log(M_*/M_{\odot}) \lesssim 10$, not presently accessible by the zCOSMOS observations, the transformation rates must drop again because the vast majority of galaxies are still in the pre-transformation state.

The effect of the environment modifying the transformation rates for star-formation activity and colour is larger than the environment's effect the morphological rates, and we conclude that the environment seems to be  more closely related to the star-formation and colour transformation processes than to the morphological transformation of galaxies. Similar conclusions have been obtained at $z \sim 0$ when using much less direct diagnostic of the environmental influence \citep[e.g.][]{Kauffmann.etal.2004, Blanton.etal.2005, Skibba.etal.2008}. For example,  \citet{Blanton.etal.2005} and \citet{Skibba.etal.2008} show that galaxies of the same colour do not show significant residual environmental dependence, while colours of galaxies of a given (similar) morphology still show environmental dependence.

This work has clearly demonstrated that, whatever physical processes are involved in the general transformations that are occuring in the galaxy population since $z \sim 1$, especially around the transition or crossing mass, these processes are evidently much more effective in groups than outside of them.  The transformation rate outside of groups is by no means zero, but might be lower than we have estimated because of residual contamination of the non-group population by group galaxies, as discussed above.

Our physical interpretation of the shorter timescales that we attribute to the colour and/or star-formation transitions, compared with those of the morphological transformations, is hindered by the uncertainty about whether 
the timescales involved are the physical timescales of the transitions (i.e. where $t_p \sim t_{\eta}$), or what we have called the statistical timescales (where $t_p \ll t_s \sim t_{\eta}$), as discussed in the Introduction.  If dealing with prompt phenomena which have the same physical timescale, e.g. say the merging of galaxies, then we would want to interpret the difference in terms of different rates of this event, e.g. different merger rates. On the other hand, in the case of longer term processes, such as strangulation, the difference would be more naturally interpreted in terms of the speed with which the given process acts.  Regardless, the fact that the transformation timescales vary rather strongly with galactic mass and with group/non-group environment, and also, but less strongly, with transition type, points towards a picture where a variety of physical processes may be in operation.

The current sampling in the 10k zCOSMOS is about $\sim 30\%$ on average, which is not sufficient to meaningfully split the population of the group galaxies into central and satellite galaxies, because two-thirds of the central galaxies will not have been observed spectroscopically. This will be much easier as the spectroscopic completeness doubles in the final 20k sample, especially if we use also the high quality photometric redshifts. Comparison with the mock group catalogue \citep{Knobel.etal.2009} suggest that both group and non group samples (defined using $M_B<-20.5-z$) are dominated by central galaxies.

\section{Conclusions}

We have used the sample of $\sim 8500$ zCOSMOS galaxies with reliable and accurate spectroscopic redshifts (uncertainty $\sim 110$ \kms) to study and compare the evolution of galaxies in $0.1<z<1$ within and outside group environments. 

We have first specifically studied the role of group environment in the morphological transformation of galaxies. We have measured the evolution of the morphological mix of galaxies (split into two broad types: early and late) for well defined sets of luminosity- and mass-complete samples of galaxies and groups.

1. In the luminosity complete samples, the fraction of early type galaxies is higher for galaxies with higher $B$-band magnitudes. There is a systematic increase in the fraction of early type galaxies in the group and field environment since $z = 1$. The build-up of the early type population depends on $M_B$ and environment: brighter early type galaxies are formed before the fainter early type galaxies, while early-type galaxies of a given $B$-band magnitude appear earlier in the group than in the field environment.

2. There is some indication within the group population
that the fraction of early types
increases with the group richness and velocity dispersion, at least at redshifts $z < 0.65$ and for $M_B<-19.5-z$.

3. Mirroring the analysis in \citet{Bolzonella.etal.2009}, the stellar mass function exhibits a different shape for samples of galaxies in different environments (groups, field and isolated) at least up to $z \sim 0.7$. The stellar mass function shows an upturn at low masses and redshifts in the group environment and more assive galaxies preferentially reside in the groups. The characteristic shape of the stellar mass functions in the group or for the isolated galaxies reflects the different relative contributions of galaxies of different types (e.g. early and late) in different environments. Due to this mass function difference, the environmental effect on galaxian properties must be examined in narrow bins of stellar mass.

4. Looking at narrow mass bins, there is again a systematic build-up of early type galaxies with stellar masses below $\sim 10^{11}$ \Msol\ since $z = 1$. At a given cosmic time, the fraction of early type galaxies in a given environment increases with stellar mass (``morphological downsizing''). For galaxies with masses below $\sim 10^{11}$ \Msol, the fraction of early type galaxies (at a given stellar mass) is always higher in the group than in the field environment, which is higher still than in isolated galaxies. Galaxies with masses above  $\sim 10^{11}$ \Msol\ show little evolution since $z = 1$, having a roughly constant fraction of early types of $\sim 0.7-0.8$ in both the group and field environments.

5. The build-up of early type galaxies, as quantified by the ``crossing mass'' (where the population is split equally between early and late type galaxies), shifts to lower masses as cosmic time passes. There is a clear environmental dependence on the rate of change of crossing mass, such that, at fixed stellar mass, the type-transformation happens first for the population of group galaxies, then for the field galaxies and finally for the sample of isolated galaxies. As time progresses, the crossing mass for morphology diverges between the three environents.  At $z \sim 0.5$, the net effect 
of the group environment is equivalent to about 0.2 dex in stellar mass, or 2 Gyr in cosmic epoch.

Broadening our analysis to compare directly the normalised transition rates in morphology with those involving galaxian colour and star-formation activity, we find the following additional points:

6. The normalised transformation rates are systematically higher in the groups than outside, by a 
factor of $\sim 2$ or, correcting for infall and the appearance of new groups, by a factor of upto $\sim 3-4$. The rates reach values, for masses around the crossing mass $10^{10.5}$ M$_{\odot}$, as high as 0.3 - 0.7 Gyr$^{-1}$ in the groups, implying transformation timescales of 1.4 - 3 Gyr, compared with less than 0.2 Gyr$^{-1}$, i.e. timescales $>$ 5 Gyr, outside of groups.

7. In the group environment the transformation rates clearly decrease with stellar mass above $10^{10.5}$ \Msol, and we infer they must also drop at the lower masses below $10^{10}$ \Msol\ that are not presently accessible by our spectroscopic data. 

8. The transformation rates are consistently higher, by about 50\%, for those transformations involving star formation and colour, than for those involving morphological changes.  Although these rate/timescale differences are not trivial to interpret (mostly because of uncertainties in the physical processes involved), they suggest faster star-formation quenching than morphological transformation, and indicate a tighter relation between the environment and star-formation related processes than with morphology-dynamical processes. 

8.  A clear and important conclusion is that, whatever their physical origin, those transformational processes which have driven the evolution of the galaxy population (around the characteristic transition or crossing mass of $10^{10.5}$ \Msol) since $z \sim 1$, occur faster, or more efficiently, in the group environment, by a factor of order 3.  Although the transformation rate is not zero outside of groups, this suggests that the group environment has played a very significant role in driving the evolution of the overall population.

The next step in the analysis will involve the forthcoming higher sampled 20k zCOSMOS catalogue.  This will not only increase the group sample by a factor of 2.5 or so, but will also allow (together with improved photo-z) a much better discrimination between the central and satellite galaxies within the groups themselves.  This will allow a more direct study of their properties (e.g. morphologies, colours, star-formation rates) which will put stronger constrain on the transformation mechanisms - both the type of processes and their timescales. Utimately, this will provide an answer whether the lifepaths of a galaxy are really different depending on if a galaxy is central or satellite \citep[e.g.][]{Bower.etal.2006, Croton.etal.2006}, or if the differences between the centrals and satellites should be more gradual \citep[e.g.][]{Simha.etal.2008}.

%\bibliographystyle{apj}
%\bibliography{refpom2}

\section{Acknowledgments}

This work has been supported in part by a grant from 
the Swiss National Science Foundation and by grant ASI/COFIS/WP3110I/026/07/0.

\end{document}